\documentclass[fleqn,usenatbib]{mnras}

\usepackage{graphicx}
\usepackage{ifthen}
\usepackage{url}
\usepackage{amsmath}
\usepackage{bm}
\usepackage{lscape}
\usepackage{rotating}
\usepackage[graphicx]{realboxes}
\usepackage{supertabular}
\usepackage{pdfpages}
\usepackage{footnote}
\usepackage{hyperref}
\usepackage{amssymb}
\makesavenoteenv{tabular}
\makesavenoteenv{table}
\usepackage{enumitem}
\usepackage[T1]{fontenc}
\usepackage{tipa}
\usepackage{accents}

\usepackage{tablefootnote}
\usepackage{caption}

\hypersetup{
  colorlinks   = true,    
  urlcolor     = blue,    
  linkcolor    = blue,    
  citecolor    = blue      
}

\usepackage{enumitem}
\setlist[itemize]{leftmargin=*} 
\setlist[enumerate]{leftmargin=*} 

\usepackage{ulem}

\def\ltsima{$\; \buildrel < \over \sim \;$}
\def\lta{\lower.5ex\hbox{\ltsima}}
\def\gtsima{$\; \buildrel > \over \sim \;$}
\def\simgt{\lower.5ex\hbox{\gtsima}}
\usepackage{amsmath}

\def\s{\ifmmode \widetilde \else \~\fi}
\def\={\overline}

\def\spose#1{\hbox to 0pt{#1\hss}}

\def\eg{{e.g.,\,}}
\def\ie{{i.e.,\,}}
\def\lta{\mathrel{\spose{\lower 3pt\hbox{$\mathchar"218$}}
     \raise 2.0pt\hbox{$\mathchar"13C$}}}
\def\gta{\mathrel{\spose{\lower 3pt\hbox{$\mathchar"218$}}
     \raise 2.0pt\hbox{$\mathchar"13E$}}}
\def\Dt{\spose{\raise 1.5ex\hbox{\hskip3pt$\mathchar"201$}}}    
\def\dt{\spose{\raise 1.0ex\hbox{\hskip2pt$\mathchar"201$}}}    

\def\dotsfill{\leaders\hbox to 1em{\hss.\hss}\hfill}

\def\FeH{{\rm[Fe/H]}}

\newcommand{\onohiula}{`\=Onohi`ula}

\loadboldmathitalic 
\title[MECHA I]{ModElling the Chemo-dynamics in Hydrodynamical cosmological Analogues (MECHA) I: The properties of simulated Local group galaxies} 
\author[F. Sestito et al.]{Federico Sestito$^{1}$\thanks{Email: \url{f.sestito@herts.ac.uk}},
Chiaki Kobayashi$^{1}$
\\
$^{1}$ Centre for Astrophysics Research, Department of Physics, Astronomy and Mathematics, University of Hertfordshire, Hatfield AL10 9AB, UK
}
\pubyear{2026}
\date{Accepted XXX. Received YYY; in original form ZZZ}

\begin{document}
\maketitle 
\label{firstpage}
\pagerange{\pageref{firstpage}--\pageref{lastpage}}

\begin{abstract}
We present the chemo-dynamical properties of dwarf satellite galaxies in our cosmological hydrodynamical zoom-in simulations of a Milky Way-type galaxy, predicting their stellar elemental abundances from hydrogen to uranium for the first time. 
We identify 223 and 118 dwarf galaxies in the high- and medium-resolution simulations, respectively, distributed along a cosmic filament centred on a Milky Way-like system.
Its satellite galaxies span wide ranges in stellar, gas, and dark matter masses.
Their star formation histories are diverse and are consistent with observations of the Local Group dwarfs; the time at which half of the stellar mass formed shows a broad distribution correlating with stellar mass, with more massive systems forming stars over extended periods. Their metallicities, velocity dispersions, luminosities, and half-light radii are in excellent agreement with observations at stellar masses $\gtrsim 10^7M_\odot$, although lower-mass simulated systems remain more metal-poor or more extended than the observed ones, as commonly found in various simulations.
We find a clear mass--metallicity relation with more massive galaxies being more metal-rich, and also age--metallicity relations of member stars with more massive systems reaching higher metallicities at earlier times. Our simulations also includes late-type metal-poor galaxies, in which metal-poor stars can also be relatively young.
Elemental abundances and metallicities possess a variety of ranges  among the simulated dwarfs. 
This is confirmed by Kolmogorov-Smirnov tests, indicating that chemical evolution histories differ significantly across systems, even at early times. 
Elemental abundance dispersions reflect the chemical evolution as a function of stellar mass: inhomogeneous mixing dominates in the lowest mass galaxies; intermediate mass systems show more homogeneous chemical evolution; 
the most massive galaxies have larger chemical dispersions likely due to chemical variety of their accreted events.
These simulations provide an ideal took for interpreting the chemo-dynamical data from upcoming large spectroscopic surveys.
\end{abstract}




\section{Introduction}
Dwarf galaxies are ideal test-bed to challenge our understanding of galactic chemical evolution, dynamical processes at various scale, and of the nature of dark matter \citep{White78,Springel06,Tolstoy09,Mcconnachie12,Weisz14,Bullock17,Simon19}. Despite of their high dark-matter fraction, dwarf galaxies possess a shallow gravitational potential, which makes them susceptible to both internal feedback processes and environmental interactions. For instance, star formation and the subsequent stellar feedback \citep[e.g.][]{ElBadry18b}, merger events \citep[\eg][]{Deason14}, ram pressure stripping \citep[\eg][]{Grebel03} and stirring \citep[\eg][]{Kazantzidis11},  reionization \citep[\eg][]{Wheeler19}, and tidal interactions \citep[\eg][]{Fattahi18} can influence 
the morphologies of dwarf galaxies \citep[e.g.][and references therein]{Higgs21,Jensen24}.

Observational studies of the Local Group dwarf galaxies have revealed a striking diversity in their star formation histories, in their gas content, in their metallicity ranges, and in their elemental abundance patterns \citep[e.g.][]{Mateo98,Tolstoy09,Kirby11,Kirby13,Venn12,Weisz14,Frebel15,Frebel16,Ji19,Taibi22,Sestito23scl,Sestito23Umi,Waller23,Sestito24Sgr2}. As summarised in \citet[][and references therein]{Kobayashi20b}, there are Local Group dwarf galaxies with a single star burst happened early on (e.g. Sculptor and Sextans), while others show signs of a less efficient and prolonged star formation (e.g. Carina and Fornax), or even with a more complex and multiple star formation episodes (e.g. Sagittarius and the Magellanic Clouds). Stars in more massive dwarf galaxies possess wide ranges of metallicities, even spanning $\sim4$~dex in [Fe/H] \citep[e.g. Sculptor,][]{Skuladottir21}, and usually their elemental abundances display signs of homogenous mixing of the interstellar medium (ISM). More massive dwarf galaxies are also chemically evolved, and the chemical pattern of their stars are affected by many nucleosynthetic channels. These nucleosynthetic channels have happened on a different time scale changing chemical compositions of gas and stars in the galaxies: from the elements produced by the explosions of the First stars; those produced in later core-collapse supernovae (SNe~II and Ibc); those processed and ejected by asymptotic giant branch stars (AGBs); and to those produced in binary systems via Type Ia supernovae (SNe~Ia) or compact object mergers.
On the other hand, stars in the less massive systems (stellar mass $\lesssim10^4\rm{M_{\odot}}$), also called ultra-faint dwarfs \citep{Simon19}, are more chemically pristine and are unlikely to have imprints of nucleosynthetic channels happened with long time scales (e.g. SNe~Ia and AGBs). The elemental abundances in ultra-faint dwarfs are usually affected by inhomogeneous mixing of the ISM during star formation, as shown by the limited high-resolution spectroscopy available \citep[e.g.][]{Ji19,Waller23,Zaremba25,Heiger26}.

From a theoretical perspective, cosmological simulations have made significant progress in reproducing the global properties of dwarf galaxies, including their stellar masses, star formation histories, and internal kinematics \citep[e.g.][]{Governato10,Hopkins14,Wetzel16,Wheeler19,Revaz18,Rey26}. However, previous cosmological simulations have very limited prediction for stellar elemental abundances, which are restricted to only a few elements. The \textsc{GEAR} simulations \citep{Revaz18}, in addition to various physical quantities, derived stellar metallicities and elemental abundances for magnesium only, with the latter quantity being over-estimated compared to observations of various Local Group galaxies (e.g. Draco, Sculptor, Sextans, Ursa Minor). Recently, \citet{Sanati23} improved the resolution of the \textsc{GEAR}, also elemental abundances of iron, magnesium and calcium, and studied the chemical properties of simulated ultra-faint dwarfs. 
Similarly, simulations from \citet{Hirai24} provided only metallicities and [$\alpha$/Fe] ratios, rather than elemental abundances. The NIHAO-UHD cosmological simulations \citep{Buck20} have been re-run by \citet{Buder25} to study the accretion history of a Milky Way (MW) analogue, including updated chemical yields but for 18 elements. 

In this paper, we use our cosmological simulations of a MW-type galaxy presented in \citet{Vincenzo20}, including chemical enrichment of all stable elements, i.e. from hydrogen to uranium, to study the properties of satellite dwarf galaxies.  The paper is divided as follows: Section~\ref{Sec:simulations} introduces the cosmological simulations and the algorithm to detect satellite galaxies. Section~\ref{Sec:properties} shows the physical properties of the simulated galaxies, such as sizes, masses, mean metallicities, velocity dispersions, and  luminosities, comparing to observations. The age--metallicity relations of stars in the simulated galaxies are discussed in Section~\ref{Sec:agemet}. The elemental stellar abundances, their dispersions, and the chemical evolution diversity within the simulated galaxies is discussed in Section~\ref{Sec:chemprop}. Conclusions are summarised in Section~\ref{Sec:conclusions}.

\section{Methods}
\subsection{Our simulations}\label{Sec:simulations}
Our simulations are run using our chemodynamical simulation code based on the \textsc{Gadget-3} \citep[see][for \textsc{Gadget-2}]{Springel05} and originally developed in \citet{Kobayashi07} to follow detailed chemical enrichment, on the DiRAC High Performance Computing Facility and also on the Linux Cluster at University of Hertfordshire in the UK.
The code includes all relevant baryon physics such as metallicity-dependent radiative cooling, star formation, stellar thermal feedback, and chemical enrichment of all stable elements from carbon to uranium, including the latest nucleosynthesis yields that can well reproduce the elemental abundances in the solar neighbourhood \citep{Kobayashi20,Kobayashi20b}.

The initial conditions are taken from the Aquarius project \citep{Springel08}, including dark matter and gas, from which stars form.
It assumes the $\Lambda$ cold dark matter cosmology, with the cosmological parameters from the one- and five-year Wilkinson Microwave Anisotropy Probe \citep{Komatsu09}, \ie  $\Omega_0 = 0.25$, $\Omega_{\Lambda} = 0.75$, $\Omega_b = 0.04$, and $H_0 = 100 \times h= 73~ \rm{km~s^{-1}~Mpc^{-1}}$. 
The Aq-C halo is used in this work as it gives rise to a MW-sized dark matter halo at the present day \citep{Scannapieco12}, and the formation processes of the thin disk was analysed in detail in \citet{Vincenzo20}.
In this paper, the high (Aq-C-4) and medium (Aq-C-5) resolution runs are used to study the properties of satellite galaxies.  Table~\ref{Table:param} reports the masses and final numbers of dark matter, gas, and star particles in the two runs.  
Note that the high-resolution run stopped at $\sim$12~Gyr, but the chemo-dynamical properties will not be affected after that time. 

In these simulations, a part of the gas particles form new star particles, once the star formation criteria  are satisfied \citep{Kobayashi04}. 
Then star particles evolve distributing their thermal, chemical and mass content into the surrounding gas particles \citep[feedback,][]{Kobayashi04}. 
The feedback and chemical enrichment depend on the local gas density as the energy and mass received by a gas particle is weighted by the smoothing kernel \citep{Kobayashi07} with 64 neighbour gas particles.
Mixing between gas particles is also included at the cooling as it affects gas cooling time-scale \citep{Haynes19}. Each stellar particle is considered as a simple stellar population, composed of stars of the same ages and metallicities \citep{Kobayashi04}, assuming the stellar initial mass function (IMF) from \citet{Kroupa02}.

Various stellar nucleosynthetic sources contribute to the chemical enrichment of gas particles from which star particles form.
The chemical enrichment includes yields from core-collapse supernovae (Type II supernovae, hypernovae \citealt{Kobayashi06}, and failed supernovae; \citealt{Kobayashi20}) with the metallicity-dependent hypernova fraction \citep{Kobayashi11b}; Type Ia supernovae \citep{Kobayashi20b} with the progenitor model in \citet{Kobayashi09}; AGB stars, electron-capture supernovae, and neutron star mergers as in \citealt{Kobayashi20}; and stellar winds of all mass-range of stars with the chemical composition at birth.

\begin{table}
    \centering
    \begin{tabular}{ccc}
        Properties & Aq-C-5 & Aq-C-4 \\ \hline\hline
        M$_{\rm{star}}$ ($10^4 \rm{M_{\odot}}$)& 13.0 & 1.63\\
        N$_{\rm{star}}$ &1,774,376 & 9,996,853   \\ \hline
        M$_{\rm{gas}}$ ($10^4 \rm{M_{\odot}}$) & 41.1 & 5.1 \\
        N$_{\rm{gas}}$ & 2,252,045 & 22,437,217  \\ \hline
        M$_{\rm{DM}}$ ($10^5 \rm{M_{\odot}}$) & 21.6 & 2.7 \\
        N$_{\rm{DM}}$ & 3,082,077 & 26,679,146  \\ \hline
    \end{tabular}
    \caption{Number of and {\it initial} masses of the stellar, gas, and dark matter (DM) particles. For the mass, values indicate the mass at the peak of the mass distributions.}
    \label{Table:param}
\end{table}

The Binary Population And Spectral Synthesis (BPASS) code\footnote{\url{https://bpass.auckland.ac.nz}} \citep{Eldridge17,Stanway18}, through the HOKI python package\footnote{\url{https://github.com/HeloiseS/hoki}} \citep{Stevance20}, is then used to derive the luminosity, and the half-light radius, in the V-band for each stellar particle. The code incorporates stellar evolution models to use the ages and metallicities of stellar particles, treated as a simple stellar population, to associate them a V-band magnitude, which is then converted into V-band luminosity.  The luminosity of a given galaxy is then the sum of member star particles' luminosities. 

\subsection{Detecting satellite galaxies in the simulations}

\begin{figure*}
\includegraphics[width=1\textwidth]{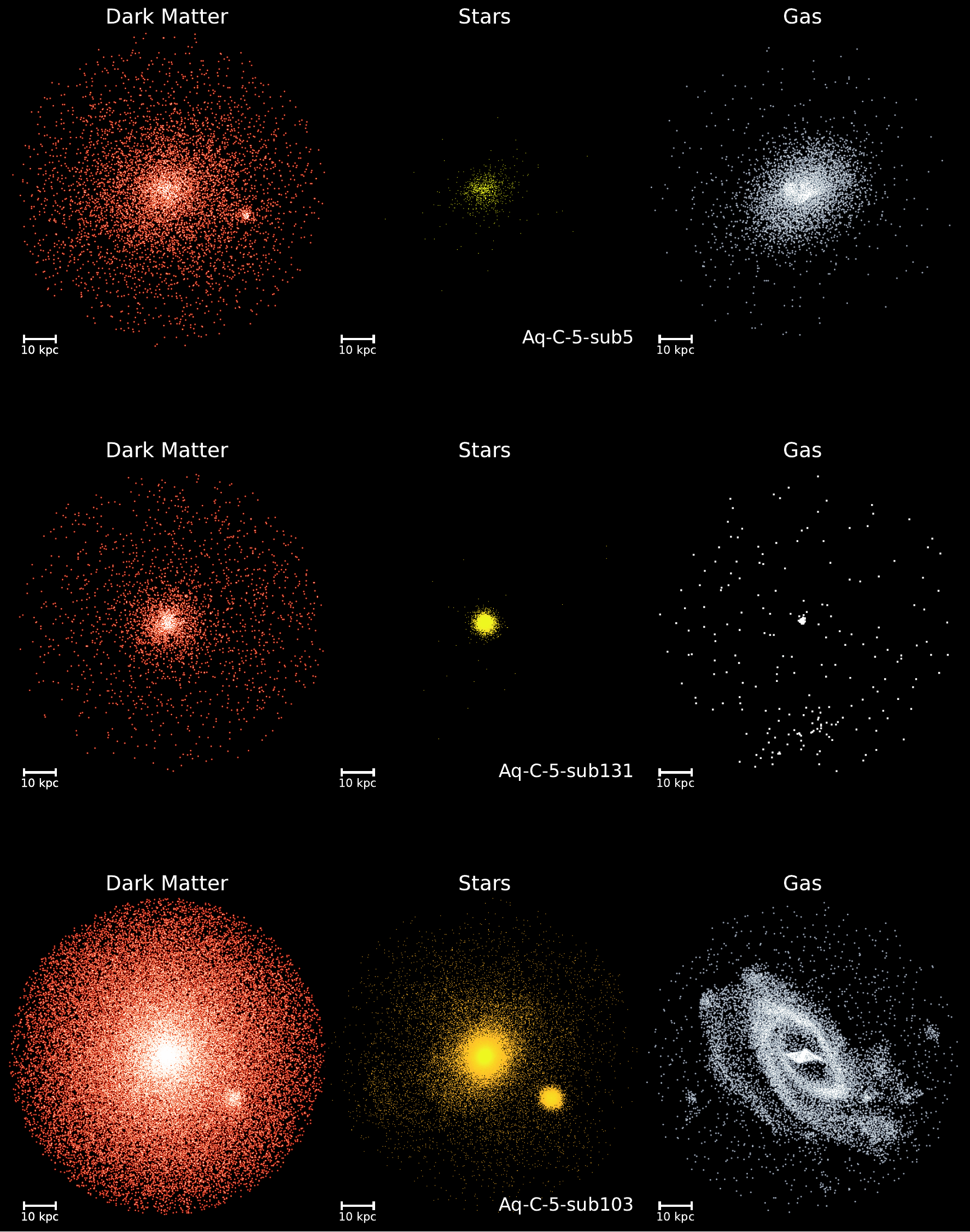}
\caption{Example of three galaxies in the medium-resolution run identified with  \textsc{HDBSCAN}. \textbf{Left panels:} dark matter particles. \textbf{Central panels:} star particles. \textbf{Right panels:} gas particles.}
\label{Fig:nice}
\end{figure*}

\begin{figure*}\ContinuedFloat
\includegraphics[width=1\textwidth]{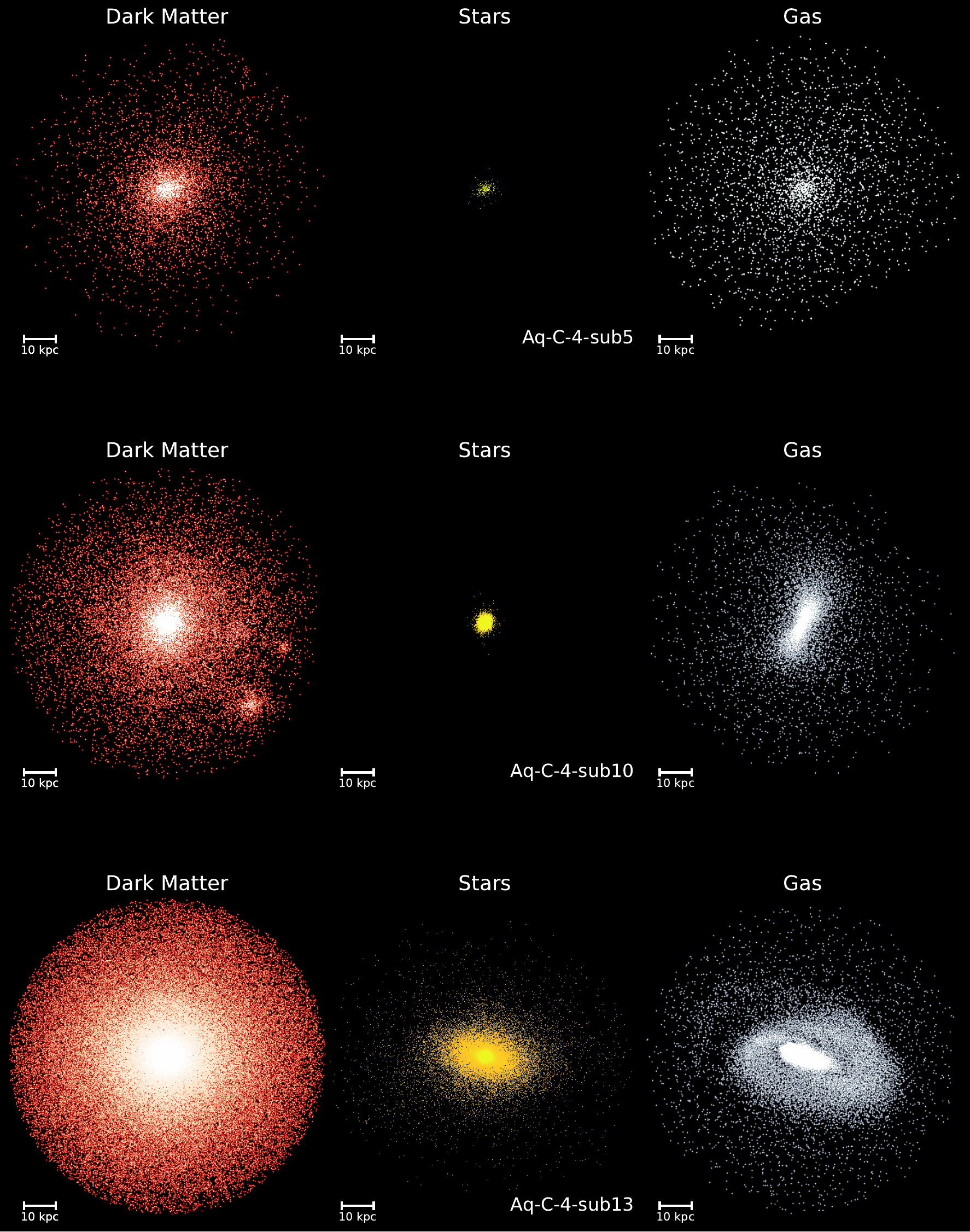}
\caption{(continued) Example of three  galaxies in the high-resolution run identified with  \textsc{HDBSCAN}. \textbf{Left panels:} dark matter particles. \textbf{Central panels:} star particles. \textbf{Right panels:} gas particles.}
\end{figure*}

We use an algorithm based on the  \textsc{HDBSCAN} python package\footnote{\url{https://github.com/scikit-learn-contrib/hdbscan}} \citep{McInnes2017} to identify spatial overdensities of stellar particles within the two simulation runs. Euclidean distances and leaf clustering algorithm are selected to better find small clusters.
A visual inspection of the dark matter, stellar and gaseous content of each structure found is performed to test whether the algorithm well separates close systems and whether the clumps are truly galaxies. In case the algorithm fails to separate two or more close systems, the algorithm is run again to improve the identification of these systems. The final selection of systems is composed of 118 and 223 simulated galaxies in the Aq-C-5 and Aq-C-4 runs, respectively, in addition to the MW-like main body.
As an example, three systems for each run are displayed in Figure~\ref{Fig:nice}, showing their dark matter (left columns), stellar population (centre columns), and gaseous content (right columns). The galaxy from Aq-C-5-sub103 (the last row) is one of the cases in which the algorithm cannot separate two interacting galaxies. In this particular case, more dispersed stellar halos are not assigned to any of the two systems. 
This slightly underestimates the galaxy masses of these cases but does not affect the conclusion of this paper.

\subsection{Centring the simulated galaxies}
\begin{figure*}
\includegraphics[width=1\textwidth]{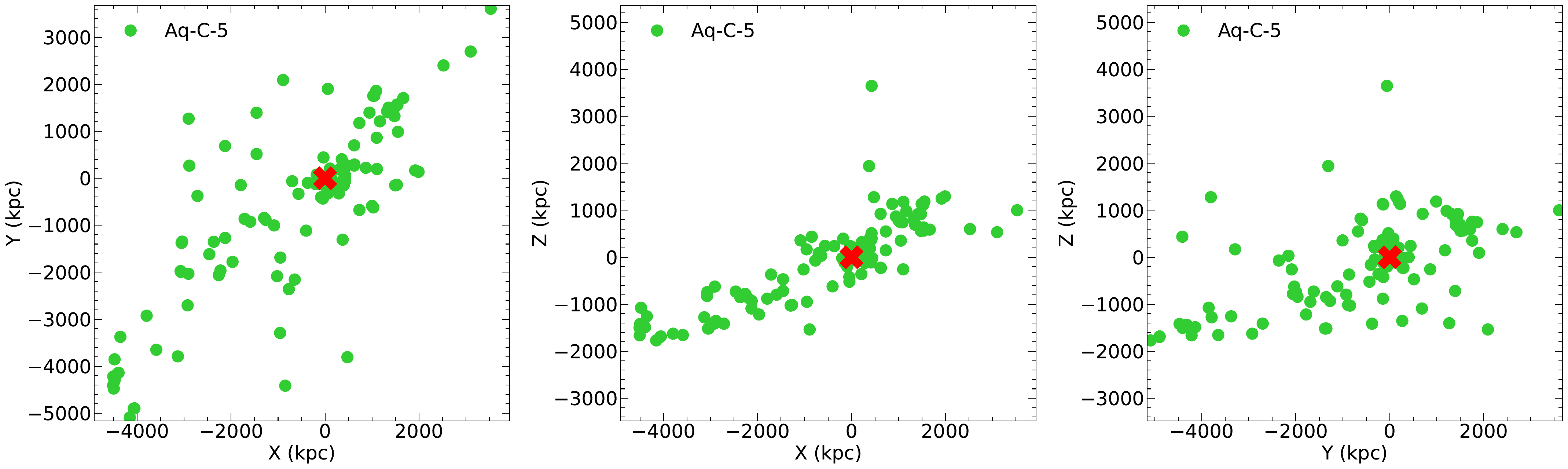}
\includegraphics[width=1\textwidth]{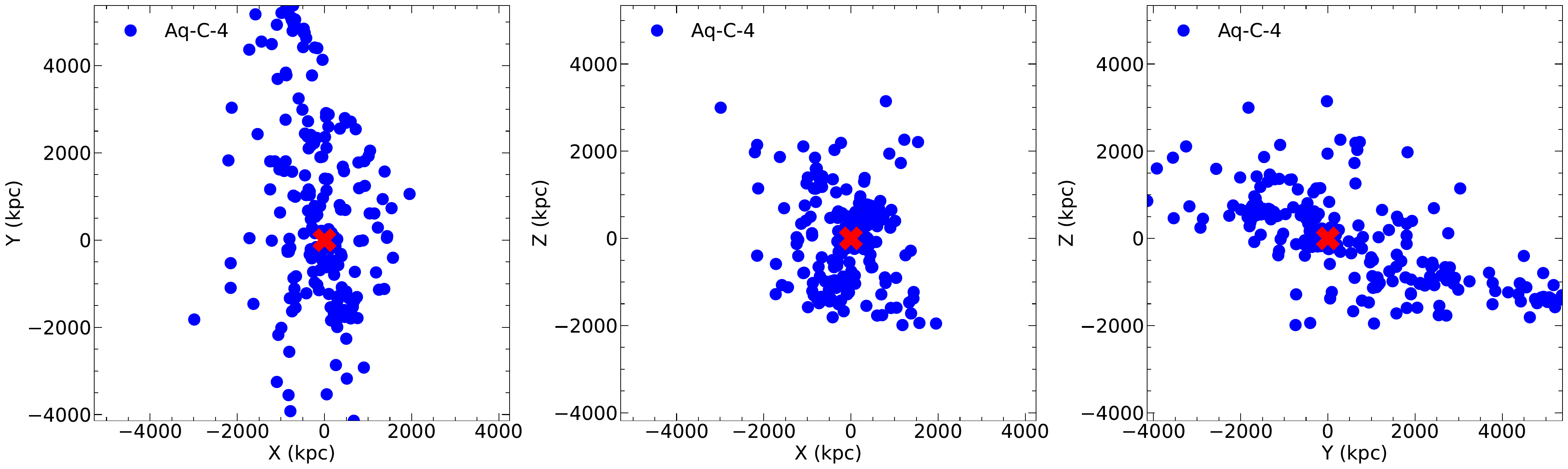}
\caption{Projected Cartesian spatial distribution of the simulated galaxies in the high (blue circles) and medium (green circles) resolution runs with respect the position of the MW-like system (red cross).}
\label{Fig:position}
\end{figure*}

The 3D positions and velocities of the simulated particles have been corrected to have the centre of the MW-like galaxy as the origin of the 3D Cartesian system of reference, by iteratively finding the centre of mass of all simulated particles (gas, dark matter, and stars). The angular momentum of the stellar particles of the central galaxy is used to align the $Z$-axis to it and  the $X$-$Y$ plane with its disc. The projected spatial distributions of the simulated galaxies in high (blue circles) and medium (green circles) resolution runs with respect of the MW-like central galaxy (red crosses) are shown in Figure~\ref{Fig:position}. The 3D distributions of the simulated galaxies are anisotropic and elongated in both runs as they follow a specific cosmic filament embedded in the cosmological initial conditions. The 3D distributions of the two runs are different mainly due to the slightly different time of the snapshots (\S \ref{Sec:simulations}).

\section{Physical properties of the simulated satellite galaxies}\label{Sec:properties}
\begin{figure*}
\includegraphics[width=1\textwidth]{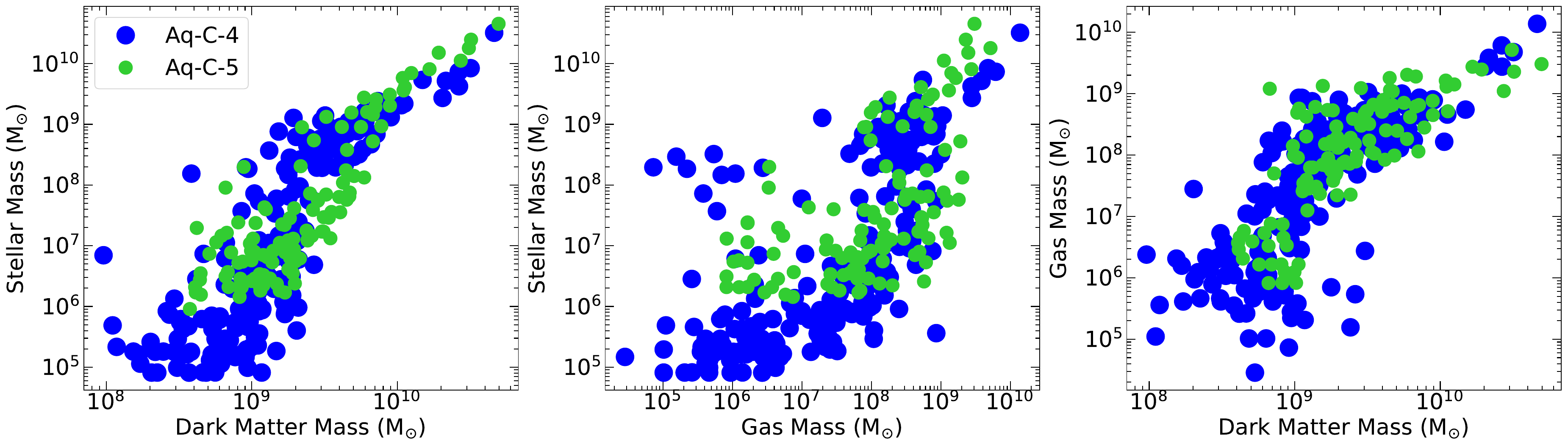}
\caption{Physical properties of the simulated galaxies. \textbf{Left:} Stellar mass vs dark matter mass. \textbf{Middle:} stellar mass vs gas mass. \textbf{Right:} gas mass vs dark matter mass. Dark matter and gas masses are measured within a sphere of 15 kpc radius of each dwarf galaxy.}
\label{Fig:mass}
\end{figure*}

Once haloes are identified, various physical quantities are derived as follows. 
Panels of Figure~\ref{Fig:mass} display the relation between the stellar mass $\rm{M_{\star}}$, the dark matter mass $\rm{M_{DM}}$, and the gas mass $\rm{M_{gas}}$ of the detected systems in high (blue circles) and medium (green circles) resolution runs. The stellar mass is simply the sum of the star particle masses found to be members of a given system. The  dark matter and gas masses are calculated by summing the masses of dark matter and gas particles within $15$~kpc from the centre of mass of all particles in each system.  While there is a clear  correlation between stellar or gas vs dark matter component for both runs (left and right panels), the relation between gas and stars is more complex (middle panel); there are various systems with relatively low stellar mass ($\rm{M_{\star}}\lesssim 10^7\rm{M_{\odot}}$) but high gas mass ($\rm{M_{gas}}\gtrsim 10^8\rm{M_{\odot}}$), resulting gas-to-star mass ratios of up to few $10^2$. Vice versa, there are also gas-poor galaxies with gas-to-star mass ratios of $\lesssim10^{-2}$. In addition, 3 and 6 galaxies in Aq-C-5 and Aq-C-4, respectively, are gas-free, not containing gas particles at all. To be noted, 2 of these gas-free galaxies have mean stellar metallicities in the very metal-poor regime ([Fe/H]~$<-2.0$) and 3 are extremely metal-poor ([Fe/H]~$<-3.0$).

\begin{figure}
\includegraphics[width=0.5\textwidth]{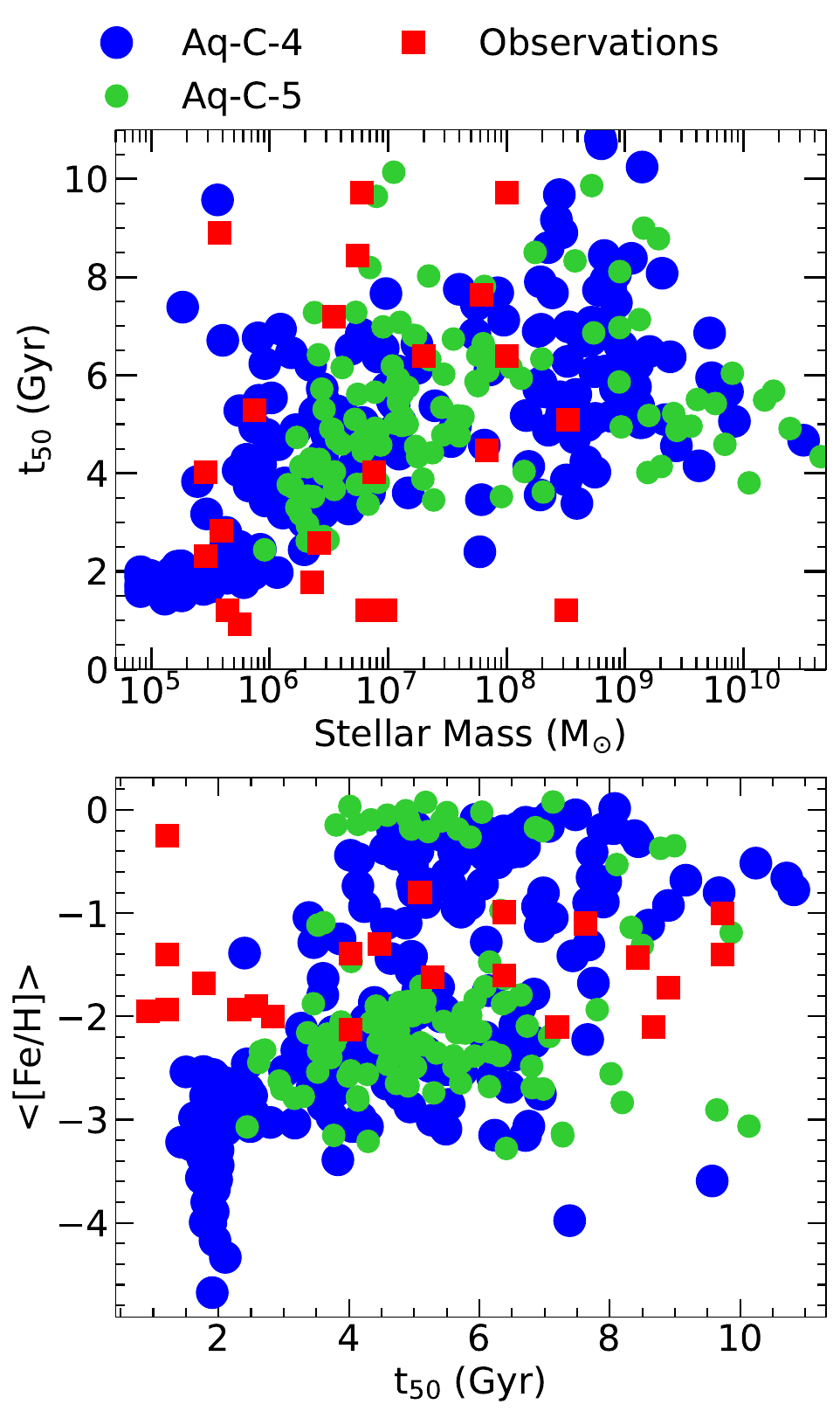}
\caption{Time at which the star formation reached 50 percent of the total, $t_{50}$ vs. stellar mass and mean metallicity. \textbf{Top:} $t_{50}$ vs stellar mass. \textbf{Bottom:} mean metallicity vs  $t_{50}$. Observations are from \citet[][and references therein; red squares]{Taibi22}.}
\label{Fig:starformation}
\end{figure}

The star formation histories of the simulated galaxies are diverse, and to quantify them, the time at which 50 percent of the stars of a given system are formed, $t_{50}$, is computed from the ages of member star particles. These $t_{50}$ are shown in Figure~\ref{Fig:starformation} as a function of the stellar mass (top) and of the mean stellar metallicity (bottom). Values from the literature compilation of Local Group galaxies by \citet[][and references therein]{Taibi22} are also plotted in Figure~\ref{Fig:starformation} (red squares).  
Our set of simulated galaxies reproduce the diversity of observed dwarf galaxies very well, covering the $t_{50}$ range of observations.
In both the simulations and in observations, there is no strong trend among the quantities displayed in  Figure~\ref{Fig:starformation}.
In the simulations, however, galaxies with low stellar masses tend to have a shorter $t_{50}$ with few exceptions. Similarly, those with short $t_{50}$ also have low mean metallicities. To be noted, there are some extremely metal-poor galaxies ([Fe/H]~$<-3.0$), in both resolutions, that spent up to 10 Gyr to form  50 percent of their stellar content. We will discuss these systems later in Section~\ref{Sec:agemet}.

\begin{figure*}
\includegraphics[width=1\textwidth]{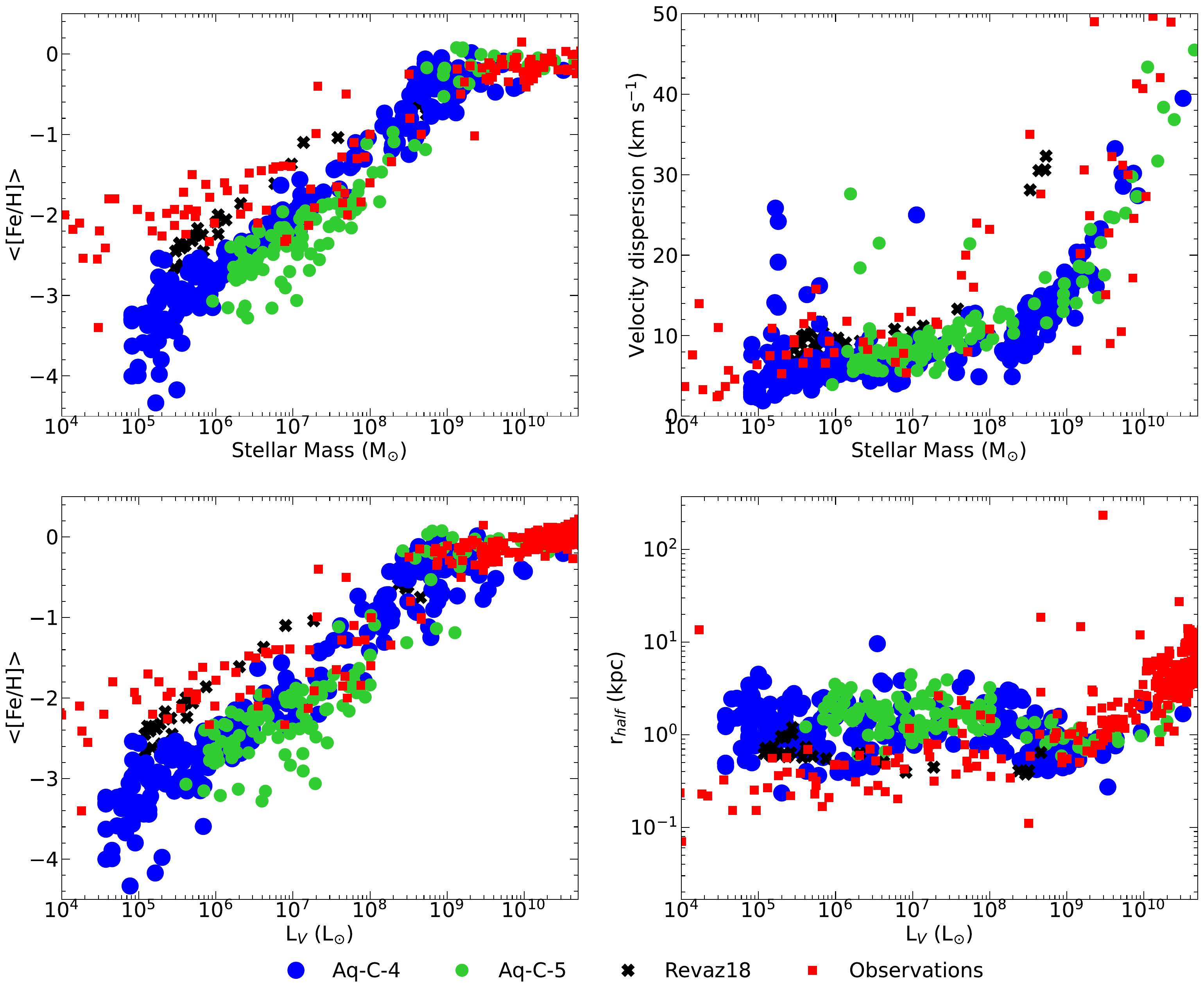}
\caption{Metallicities, velocity dispersions, half-light radii, and luminosities of the simulated galaxies. \textbf{Top left:} mean metallicity vs stellar mass.  \textbf{Top right:} velocity dispersion vs stellar mass. \textbf{Bottom left:} mean metallicity vs luminosity in the V-band.  \textbf{Bottom right:} half-light radii vs  luminosity in the V-band. Simulated galaxies from \citet{Revaz18} are displayed with black crosses. The compilation of the observed Local Group galaxies (red squares) are from \citet{Mcconnachie12}, \citet{Yuan25} for the most metal-poor system (C--19), and from the SDSS MaNGA-Pipe3D dataproduct \citep{York00,Lacerda22,Sanchez22}.}
\label{Fig:various}
\end{figure*}

Figure~\ref{Fig:various} displays the relations between the mean stellar metallicities <[Fe/H]>, the stellar masses, the (total) velocity dispersions, the V-band luminosities, and the half-light radii of the simulated galaxies (blue and green circles), comparing to the observed galaxies in the Local Group (red squares) and other simulated galaxies in the GEAR hydrodynamical cosmological simulations \citep[black crosses,][]{Revaz18}. The properties of the observed galaxies are taken from various literature: \citet{Mcconnachie12} for the most close-by galaxies, which have  $\rm{M_{\star}}\lesssim3\times10^9\rm{M_{\odot}}$; the SDSS survey MaNGA-Pipe3D dataproduct\footnote{\url{https://skyserver.sdss.org/dr17/MoreTools/browser?table=mangaPipe3D}} \citep{York00,Lacerda22,Sanchez22} for the farthest and more massive galaxies ($\rm{M_{\star}}\gtrsim10^9\rm{M_{\odot}}$); and \citet{Yuan25} for the most metal-poor system, C--19 \citep[see also][]{Martin22,Errani22b,Yuan22,Bonifacio25,Venn26}. For the most close-by galaxies, metallicities and velocity dispersions are obtained from spectroscopy, while luminosities and stellar masses are usually derived assuming a given stellar population model for the small  number of observed red giants \citep[see references in][]{Mcconnachie12}. The SDSS galaxies are selected to be likely elliptical galaxies ($\rm{P(E) > 0.7}$), with good quality observations (QCFLAG~$=0$). The velocity dispersions in SDSS are measured at the half-light radius, while the metallicities are weighted by the mass of the stellar populations. The luminosities of the SDSS galaxies are derived converting their absolute V-band magnitudes.

The mean metallicity vs stellar mass relation panel (top left) shows very good agreement between our simulations and observations at $>10^7M_\odot$.
At the lower mass, observed metallicities become roughly constant and are higher than those of simulated galaxies.
The lack of ``metallicity floor''  at low mass is common across various cosmological simulations suites \citep[e.g.][]{Wetzel16,Maccio17,Revaz18}.
The lower systemic metallicities was linked to the assumptions on the IMF, which is considered constant in these simulations \citep[e.g.][]{Revaz18}. A more in-depth model should consider a different  IMF for Population~III stars \citep[e.g.][]{Bromm13}, which might affect the mean metallicities more at the lowest mass systems. Alternatively,  the enrichment algorithm might not be able to cope with very rapid and very small star formation bursts \citep[e.g.][]{Maccio17}; or the numerical resolution might be too low to reproduce the observed properties \citep[e.g.][]{Wetzel16}. More recently, \citet{Rey26} succeeded in reproducing the metallicity plateau with simulations that include a top-heavy IMF, favouring pair-instability supernovae with a higher Fe yield per explosion, and higher mass haloes able to retain more efficiently metals.

However, our simulations reach the lower metallicity range of the extremely metal-poor galaxy recently discovered. C--19 is the system that  mitigates the value in our simulations with those of the observed Local Group at the lowest metallicity and stellar mass. C--19 is a stellar stream in the Milky Way and, given its small metallicity dispersion, so far, it was associated with a disrupted globular cluster \citep{Martin22,Yuan22,Bonifacio25}. However, from its dynamical properties, it is not clear whether the system was embedded in a dark matter halo at the formation time \citep{Errani22b,Carlberg25}. Recently, \citet{Venn26} analysed high-resolution spectroscopic observations and suggested that star formation in C--19 was extremely rapid and efficient, and occurred in a single burst. They found neither evidence of inhomogeneous mixing contrary to what is observed in other ultra-faint dwarf galaxies \citep[e.g.][]{Simon19},  nor indication of multiple stellar populations unlike  globular clusters \citep[e.g.][]{Bastian18}. The agreement to our simulations might support that C-19 is a galaxy.

The velocity dispersions of the simulated galaxies, as a function of the stellar mass (top right panel), very well match the distribution observed in Local Group galaxies. On the observational side, the velocity dispersions of systems with $\sim 2\times 10^7 < \rm{M_{\star} (M_{\odot})} < 10^{10}$ is very scattered, showing a less pronounced trend with the stellar mass. The systems in this mass range and with large velocities include satellites of the Andromeda galaxy (NGC~147, NGC~185, and NGC~205), irregular dwarfs that might also be tidally perturbed (the Small Magellanic Cloud,  NGC~6822, and the Wolf–Lundmark–Melotte galaxy), and the Milky Way tidally perturbed satellites Canis Major and Sagittarius. This is consistent with our simulations where we find that the galaxies with relatively large velocity dispersions are perturbed by  interactions with other systems when we look at their histories.

The bottom left panel in Figure~\ref{Fig:various} shows the mean metallicity vs V-band luminosity. Our simulations very well match the luminosity of the observed galaxies, with the exceptions of those at the lowest metallicities, which reflects the fact that our simulations have lower metallicities than those of the Local Group systems, as previously discussed for the top left panel.

The bottom right panel in Figure~\ref{Fig:various} displays the half-light radius as a function of the luminosity of our simulations, as well as those for the observed galaxies. There is a general agreement between observations and simulations, although the least massive simulated galaxies have slightly larger half-light radius than the observed counterparts. The mismatch at the lowest mass regime might be due to the numerical resolution  \citep[see][for further discussion]{SantosSantos26}. 
However, we should note that our higher-resolution run (blue) show only a slight improvement compared with our lower-resolution run (green), which indicates that the mismatches in Figure~\ref{Fig:various} are caused by the assumptions in astrophysics.

\section{The age-metallicity relations}\label{Sec:agemet}
\begin{figure*}
\includegraphics[width=1\textwidth]{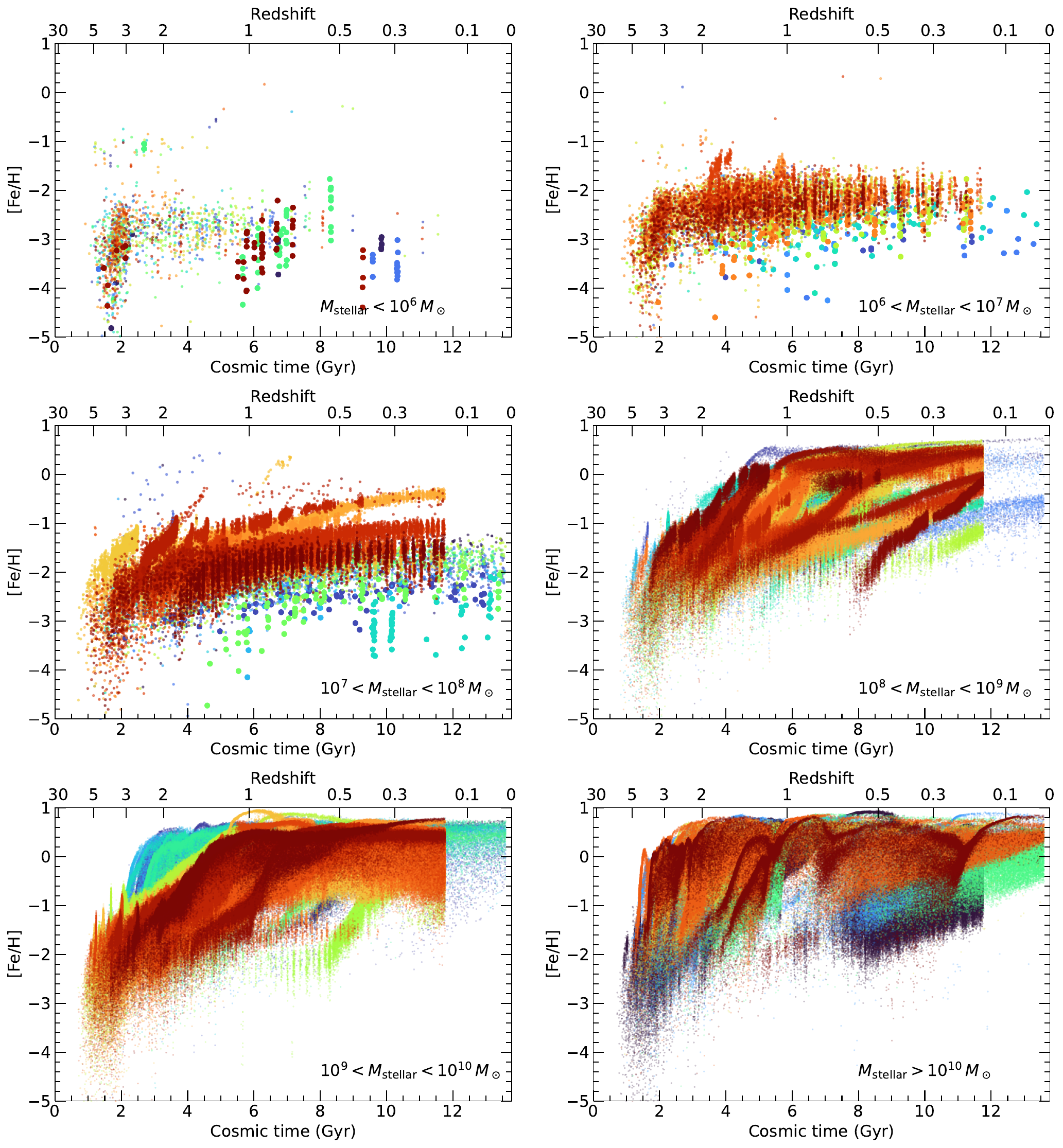}
\caption{Stellar metallicities as a function of Cosmic time for all galaxies in the Aq-C-4 and Aq-C-5 simulations. Each panel corresponds to a different stellar mass bin. Inside a given mass bin, stars of the same galaxy have the same colour. Larger markers denote galaxies with mean metallicity $\FeH\leq-2.5$ and with a late star formation, $t_{50} > 6$~Gyr and they are present in the three least massive  bins ($\rm{M_{stellar} <10^8M_{\odot}}$). The size and transparency of the markers in the three most massive bins are reduced for a better visualisation. The Aq-C-4 run ends at $\sim$12 Gyr.}
\label{Fig:agemet}
\end{figure*}

Although it is very difficult to estimate ages of stars in Local Group dwarf galaxies, stellar particles in our simulations have stellar ages, which allows us to investigate enrichment histories in detail. The formation time of the stellar particles (in cosmic time and cosmological redshift) vs their metallicities, of all galaxies in the Aq-C-4 and Aq-C-5 runs, are shown in the six panels of Figure~\ref{Fig:agemet}. Systems are placed in the panels of Figure~\ref{Fig:agemet} according to their total stellar mass. Each colour in a given panel denotes stellar members of the same galaxy. Systems with mean $\FeH<-2.5$ and with late star formation ($t_{50} >6$~Gyr) are denoted with larger markers. The overall picture is that the metallicity in a given system increases as a function of cosmic time, as also found and expected by various  cosmological simulations \citep[\eg][]{ElBadry18,Starkenburg17a,Buck20,Sestito21}. 

The top left panel ($\rm{M_{\star}<10^6\ M_{\odot}}$ contains the least massive systems, which also have mean metallicities $\FeH \lesssim -2.5$, i.e. lower than those of the systems in the other bins.
As the total stellar mass bin increases, the systems' mean metallicities increase (as also discussed for Figure~\ref{Fig:various}) and the shape of the age-metallicity relations becomes more complex. For instance, more massive galaxies are more efficient in retaining the metals, hence reaching higher metallicities in a shorter Cosmic time. Systems with $\rm{M_{\star}} \lesssim 10^8 \rm{M_{\odot}}$ (first to third bins) does not reach [Fe/H]~$\sim0.0$, with some exceptions. Systems with mass range $10^8 <  \rm{M_{\star}} < 10^9 \rm{M_{\odot}}$ (fourth bin) have spent at least $\sim$4.5~Gyr to produce stars with solar- or super-solar [Fe/H]; this time can be as short as $\sim$1.8~Gyr for the most massive galaxies ($\rm{M_{\star}} \gtrsim 10^{10} \rm{M_{\odot}}$, the last bin).
In the more massive and extended systems, the various complex features of this space are attributed to two main processes: i) the dilution of the interstellar medium with pockets of pristine gas, which form  more metal-poor younger stars, or ii) the merging of galaxies with  different chemical histories.

As noted  for Figure~\ref{Fig:starformation}, there are some very and  extremely metal-poor systems that formed stars only later on, having  $t_{50}$ up to $\sim10$~Gyr. There are 18 systems with mean $\FeH<-2.5$ and  $t_{50}>6$~Gyr, and they are shown in Figure~\ref{Fig:agemet} with larger markers. All the late-type metal-poor galaxies have stellar masses $M_\star\lesssim10^{8}\rm{M_{\odot}}$. Some of them has a late and continuous star formation, while other galaxies are bursty, forming stars in multiple bursts, even separated by several Gyr.
The finding of such young and metal-poor galaxies is very surprising as the common picture is that the most metal-poor stars should also be very ancient, which is the idea at the base of Galactic Archaeology. I.e. the Galactic  metal-poor stars have been used to trace the assembly history of the Milky Way, and Local Group satellites are assumed to be very old. 
In our simulations, however, some of these metal-poor systems are also gas-rich. More detailed nature of these extremely metal-poor young systems will be reported in Sestito~\&~Kobayashi (in prep.).

\section{Chemical properties of the systems}\label{Sec:chemprop}

\subsection{Elemental abundance diagrams}

\begin{figure*}
\includegraphics[width=1\textwidth]{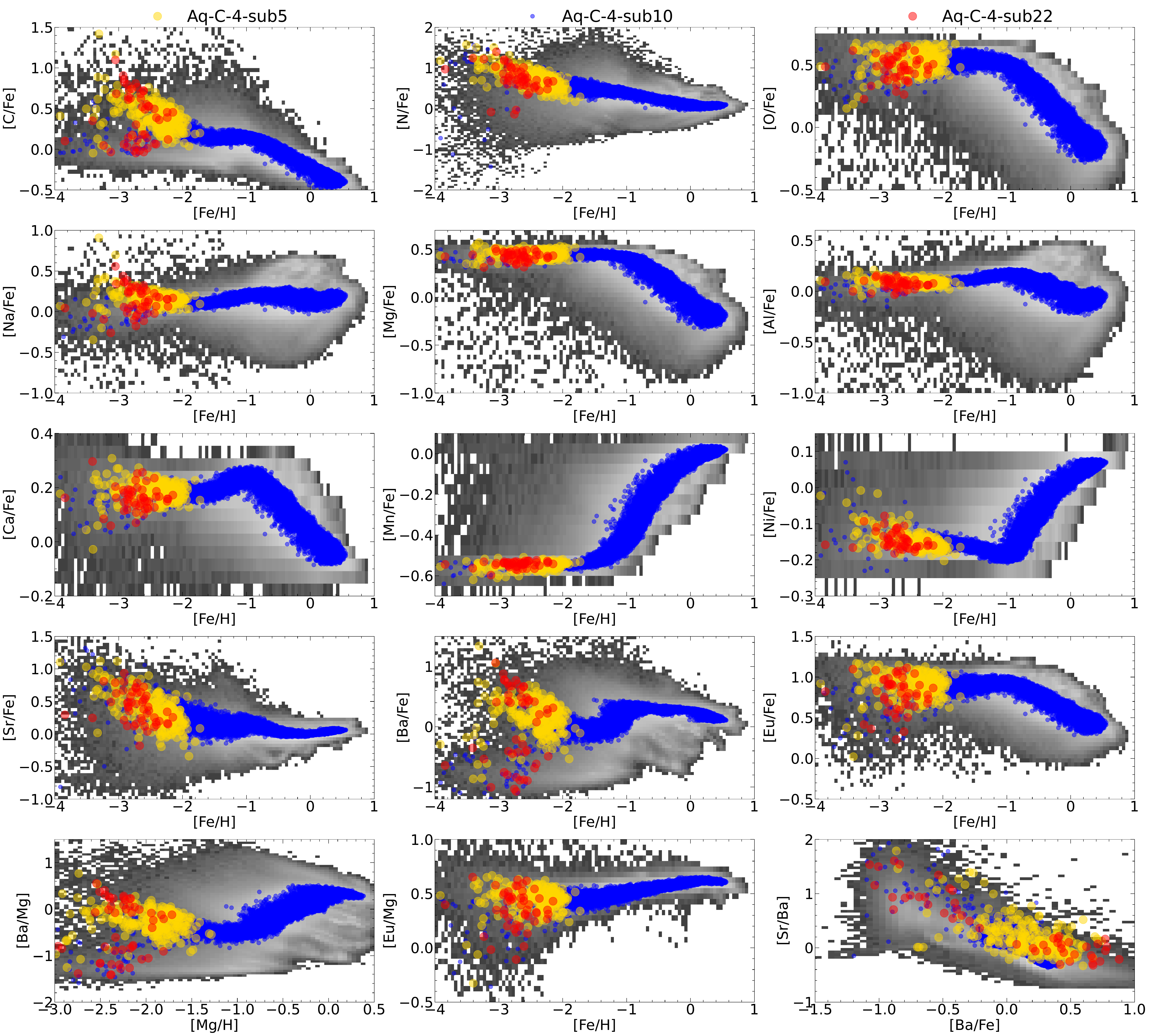}
\caption{Chemical abundance ratios as a function of metallicity for the MW-like system and 3 simulated dwarf galaxies. The number density of stars in the MW-like galaxy is shown with a grey colour-map, with lighter colours for the higher concentration of  star in a given pixel. Pixels have size 0.05 by 0.05 dex. The three simulated galaxies have stellar masses of $\rm{M_{\star}} = 8.33\times10^{5} \rm{M_{\odot}}$ (red, Aq-C-4-sub22), $\rm{M_{\star}} = 4.18\times10^{6} \rm{M_{\odot}}$  (gold, Aq-C-4-sub5), and $\rm{M_{\star}} = 6.69\times10^{8} \rm{M_{\odot}}$ (blue, Aq-C-4-sub10).}
\label{Fig:chemical}
\end{figure*}

Our simulations include detailed chemical evolution, thus each stellar particle  has elemental abundances for all the elements up to U.
The 15 panels of Figure~\ref{Fig:chemical} display the elemental abundance ratios for three example systems (red, gold and blue) and for the  MW-like galaxy (grey density pixels) in the high-resolution run. The top 4 rows  show the [X/Fe] ratios as a function of the [Fe/H] of 2 light elements (C, N), 3 $\alpha$-elements (O, Mg, and Ca), 2 odd-$Z$ elements (Na, Al), 2 Fe-peak elements (Mn and Ni), and 3 neutron-capture elements (Sr, Ba and Eu). The bottom row displays key combination of the abundance ratios, [Ba/Mg] vs [Mg/H], [Eu/Mg] vs [Fe/H], and [Sr/Ba] vs [Ba/Fe].

As thoroughly discussed in \citet{Kobayashi20} and \citet{Kobayashi25}, the contribution of asymptotic giant branch stars (AGBs) in the production of C and N reaches 49 and 74 percent, respectively at a cosmic time of $\sim$9~Gyr. The rest of C and N is produced by core-collapse (Type II and Ibc) supernovae (CCSNe).
The $\alpha$-elements (here O, Mg, and Ca), Na, and Al are produced mostly by CCSNe. 
At later times, Type Ia supernovae (SNe~Ia) are also responsible for the Ca production. The Fe-peak elements (Fe, Ni and Mn) are produced mainly from CCSNe and SNe~Ia, with the latter being the predominant production site. The neutron-capture elements (here Sr, Ba and Eu)  have multiple nucleosynthetic pathways, i.e. CCSNe, AGBs and compact object mergers (a neutron star, and another neutron star or a black hole), and it is still an open question which of them are the most predominant sites, also as a function of the metallicity or cosmic time \citep[][and references therein]{Kobayashi23,Kobayashi25}.

In the figure, the three example systems are chosen to span a wide range of stellar masses. The least massive system of the three, Aq-C-4-sub22 (red),  has a stellar mass of $\rm{M_{\star}} = 8.33\times10^{5} \rm{M_{\odot}}$, and all of its stars have [Fe/H]~$<-2.15$.  Aq-C-4-sub5 (gold), also displayed in Figure~\ref{Fig:nice},  has a stellar mass of $\rm{M_{\star}} = 4.18\times10^{6} \rm{M_{\odot}}$, and consists of stars with [Fe/H]~$<-1.7$. 
More massive system, Aq-C-4-sub10 (blue), also showed in Figure~\ref{Fig:nice},  has a stellar mass of $\rm{M_{\star}} = 6.69\times10^{8} \rm{M_{\odot}}$, and the metallicity of its  stars reaches  [Fe/H]~$\sim+0.5$.

The two low-mass systems shows signs of inhomogeneous enrichment; this effect  is visible in the elemental dispersion of most of the elements, with the exception of Mg, Al, and Mn. Conversely, Aq-C-4-sub10 underwent more homogeneous chemical evolution.  
This relatively massive system also has contribution from SNe~Ia at [Fe/H]~$\gtrsim-1.0$. As SNe~Ia  produce much more Fe than other sources, the effect of SNe~Ia is seen as a net decrease in most of the [X/Fe], with the exception of Mn and Ni. These Fe-peak elements are also largely produced by SNe~Ia, which appears as an increase of their [X/Fe] ratios. As noted previously, Ca is also produced by SNe~Ia but with a smaller amount compared to Fe, which explain the decrease of [Ca/Fe]. The neutron-capture process displays a wider range of values, i.e. higher dispersion, even in the most massive galaxy in Figure~\ref{Fig:chemical}. In our simulations, [Sr/Fe], [Ba/Fe], and [Eu/Fe] vary up to $\sim1.0$, $\sim2.0$, and $\sim1.5$ dex, respectively, even among stars at similar metallicities and in the same galaxy.

The bottom row of Figure~\ref{Fig:chemical} shows the key combination of abundance ratios that have been used in observational works \citep[e.g.][]{Mashonkina17b,Skuladottir19,Sestito24Sgr} to determine whether a given galaxy has contributions from a variety of channels, namely for the s-process (bottom left), for the r-process (bottom centre), and extra neutron-capture processes (bottom right). As Mg is mostly constant and mostly produced by CCSNe, it can be used to test whether other nucleosynthetic channels produce Ba and Eu or not.

In our simulations, the large elemental abundance dispersion at low metallicities, as well as the clear trends at high metallicities, in the  [Ba/Mg] vs [Mg/H] diagram (bottom left) indeed predict the presence of additional, delayed channels in the Ba production i.e. AGBs \citep[e.g.][and references therein]{Skuladottir19}. 
As not much Eu is  not produced by the s-process, the increasing [Eu/Mg] ratio as a function of \FeH{} (bottom centre) indicates that Eu is also produced in delayed sources, such as compact object mergers \citep[e.g.][and references therein]{McWilliam13,Skuladottir19}. 
These trends are much more clearly seen for the simulated dwarf galaxies than the simulated MW-like galaxy (gray colour map).

The [Sr/Ba] vs [Ba/Fe] diagram (bottom right) shows that, even in the same system, the [Sr/Ba] and [Ba/Fe] ratios span in a wide range.  
The solar or sub-solar [Sr/Ba] ratios at high [Ba/Fe] indicates Sr is also produced by the s-process, while the super-solar [Sr/Ba] ratios at low [Ba/Fe] suggests Sr is produced by additional sources other than the s-process or the r-process \citep[e.g.][]{Mashonkina17b}.

Our simulations includes the same chemical evolution model as in \citep{Kobayashi20}.
In addition to the s-process in AGB stars and the r-process in neutron-star mergers \citep{Wanajo14} and magneto-rotational supernovae \citep{Nishimura15}, the weak-r process in electron-capture supernovae \citep{Wanajo13} are also included. The large variation in [Sr/Ba] is caused by the narrow mass range and the small ejecta mass.
Other neutron-capture processes that are not included in our simulations are: 
the weak s-process in the He-burning phase of massive stars \citep{Raiteri91}; charged particle reactions in CCSNe \citep{Woosley92}; the weak s-processes in rotating  massive stars \citep{Pignatari08,Limongi18}; weak r-process from hypernovae \citep{Izutani09}; and neutrino-driven winds \citep{Arcones07,Bliss17}.
Such large scatter in the [Sr/Ba] ratio is also observed in the Milky Way and in Local Group dwarf galaxies \citep[e.g.][and references therein]{Mashonkina17b,Ji19,Sestito24Sgr}. However, some of the observed dwarf galaxies, i.e. Gru~I, Bootes~1 and Sculptor, can populate the part of the diagram with  [Sr/Ba]~$\lesssim-0.5$ and [Ba/Fe]~$\lesssim-0.5$ \citep{Feltzing09,Norris10,Gilmore13,Ishigaki14,Frebel16,Mashonkina17b,Ji19}.

\subsection{Quantifying elemental abundance dispersions}

\begin{figure*}
\includegraphics[width=1\textwidth]{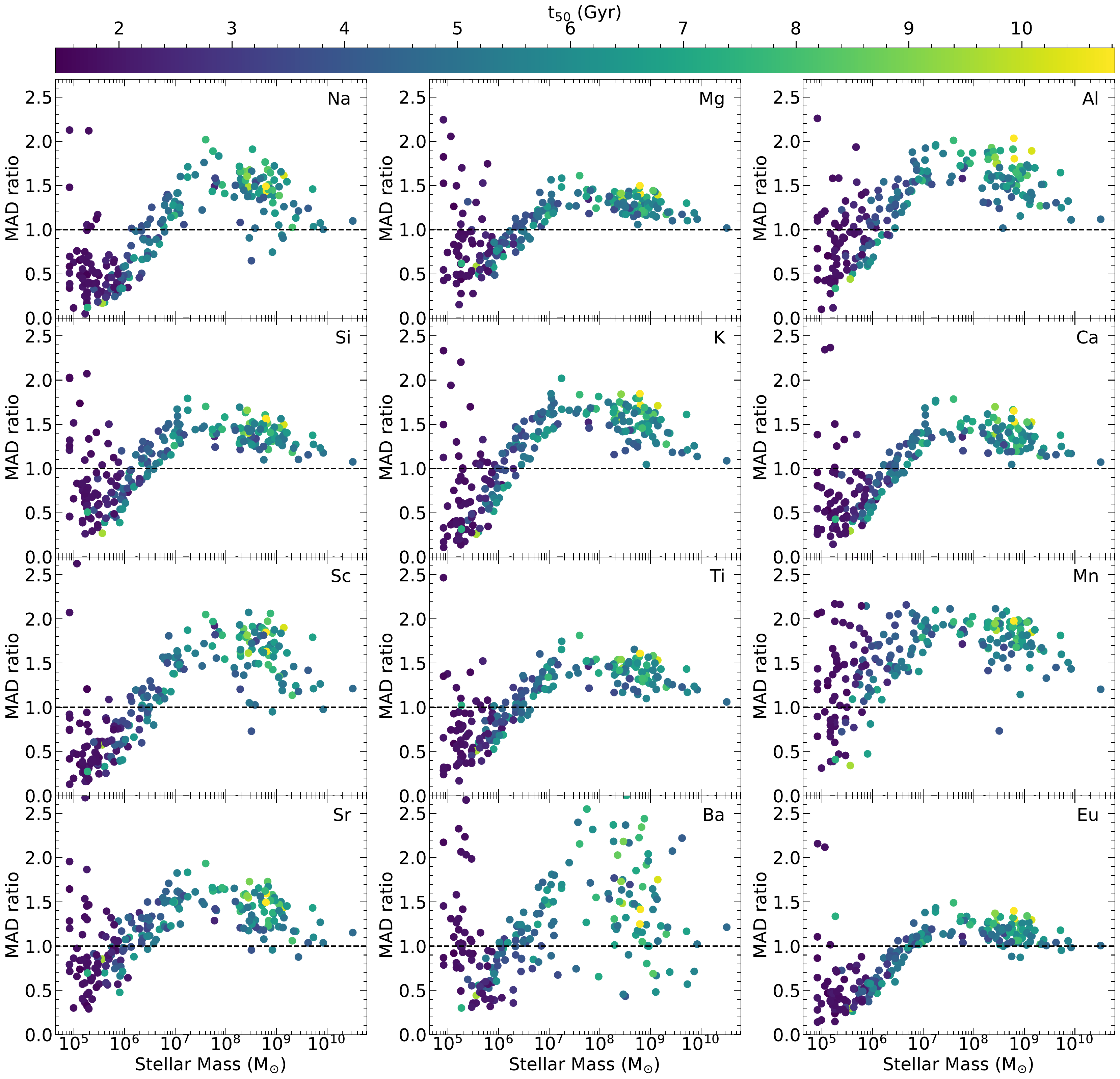}
\caption{MAD ratios (MW/dwarf) vs stellar mass for the high-resolution run, colour-coded by $t_{50}$. Each panel represent a different atomic element. The MAD ratios are calculated by dividing the MAD of [X/Fe] in the MW-like system by the MAD of [X/Fe] in a given dwarf, in both cases selecting stellar particles with [Fe/H]~$<-2.0$.}
\label{Fig:madratio}
\end{figure*}

Figure~\ref{Fig:chemical} demonstrates that the simulated galaxies  have a variegated chemical evolution with some elements displaying a narrower distributions than others. In this section, we quantify how the elemental abundance dispersions vary as a function of the stellar masses also with respect of the Milky Way-like system. The simulated Milky Way-like galaxy (grey colour map in Figures~\ref{Fig:chemical}) shows larger ranges of the [X/Fe] compared to its satellite galaxies, because the central galaxy has a mixture of stars born in-situ and accreted.

We select star particles with $\FeH<-2.0$ in each satellite galaxy as well as the central galaxy. For each [X/Fe], the median absolute deviation (MAD) in the simulated Milky Way is calculated, which is compared with the MAD of satellite galaxies. Each panel of Figure~\ref{Fig:madratio} displays the MAD ratios (MW/dwarf) of a given [X/Fe] as a function of the  stellar masses of galaxies. Markers are colour-coded by the $t_{50}$ quantity (see Section~\ref{Sec:properties}). Elements are chosen to have relatively strong spectral-line features in the optical wavelengths for metal-poor stars, therefore, are observable with most of the ground-based high-resolution spectrographs.

In this MAD ratio space,  values $>1$ ($<1$) mean that the given elemental abundance dispersion  is larger (smaller) in the simulated Milky Way than in the satellite galaxy. 
In Figure~\ref{Fig:madratio}, a very similar  trend is seen for many [X/Fe] panels, but with different MAD ratio values or slopes. For most of the least-massive systems ($M_\star\lesssim10^{6}\rm{M_{\odot}}$), the MAD ratios are $<1$, with the exception of a small number of systems at the low-mass end. The MAD ratios increase with the total stellar mass, reaching a value close to $\sim1$ around $M_\star\sim10^{6}\rm{M_{\odot}}$. At $10^{7} \lesssim M_{\star}/\rm{M_{\odot}} \lesssim 10^8$, the MAD ratios reach their maximum, with values $>>1$. For the most massive galaxies ($M_\star\gtrsim10^{8}\rm{M_{\odot}}$), the MAD ratios start to decrease, re-approaching  values to $\sim1$.

This complex trend common among all the elements can be be explained by three enrichment processes dominant at different mass ranges. 
The majority of the least-massive galaxies are affected by {\it inhomogeneous mixing}, which increases elemental abundance dispersions (see an example in Figure~\ref{Fig:chemical}). It reflects into the MAD ratios, decreasing the values even smaller than 1. In slightly larger and more massive galaxies, the star forming gas is retained inside the galaxy and the chemical elements produced by the stars are recycled, leading to more  {\it homogeneous chemical evolution}. In this case, galaxies have small elemental abundance dispersions. Compared with the simulated MW, the MAD ratios become larger than 1. Lastly, the most massive galaxies are also experienced a prolonged {\it hierarchical accretion} phase, having metal-poor population coming from both in-situ and accreted systems, similar to the simulated MW. This results in large dispersions in [X/Fe] and lower values of the MAD ratios.

Recently, \citet{Sestito26} used this MAD ratio space to discuss whether a sample of very metal-poor stars ($\FeH\leq-2.0$) with high-eccentric orbits and confined to the MW disc resembles dwarf galaxies or the MW Galaxy. They found that this peculiar ``planar'' population has MAD ratios $>1$ for various [X/Fe], meaning elemental abundance dispersions much smaller than in the Galactic halo. Instead, the elemental dispersions are similar to those of classical dwarfs. One scenario that might explain the chemo-dynamical properties of this ``planar'' sample is that these stars might have undergone a homogenous chemical evolution and they might have originated in the same formation site, which they dubbed \textit{Loki} \citep{Sestito26}. If that is truly the case, our findings are in agreement with their results; the baryonic mass provided in \citet{Sestito26} was in the range of stellar masses where we see MAD ratios well above $1$, due to homogeneous chemical evolution. The MAD ratio space can be an effective diagnostic to investigate the nature of accreted populations found through dynamical tagging. 
A good practice in this exercise would be to select those elements that present  constant [X/Fe] ratios as a function of \FeH{}, i.e.  $\alpha$ and Fe-peak elements at \FeH{} before SNe~Ia are start to contribute.

There is no clear trend of the MAD ratios with the $t_{50}$ quantity, however at the lower-mass end ($\rm{M_{\star}} \lesssim 10^7 \rm{M_{\odot}}$), those galaxies with higher $t_{50}$ have also lower MAD ratios compared to systems with similar masses. This indicates that even such low-mass systems might experience accretion events (which will be shown in Sestito~\&~Kobayashi, in prep.). 
The panels of Figure~\ref{Fig:madratio} are also informative on which elements are the most useful to identify enrichment processes. Those elements that show a larger scatter in their MAD ratios, e.g. Ba, are produced by both s- and r-processes at different time scale, and should be avoided for this exercise.

\subsection{Is early chemical evolution a universal process?}

\begin{figure}
\includegraphics[width=0.5\textwidth]{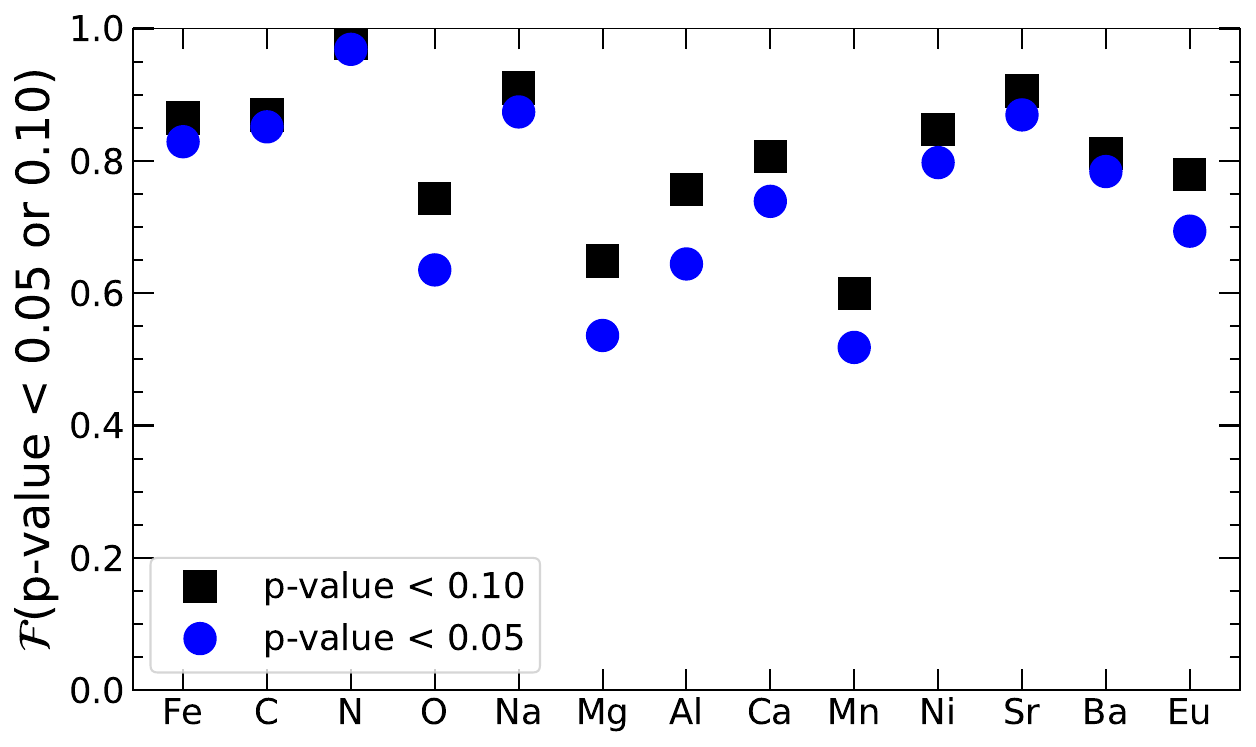}
\caption{Fraction of the number of galaxies for which the Kolmogorov-Smirnov test can be rejected for [Fe/H] and 12 [X/Fe] ratios. Blue and black circles denote the fraction for which the K-S test is rejected adopting a p-value threshold of  $< 0.05$ (blue circles)  and $<0.10$ (black squares).}
\label{Fig:kolmogorov}
\end{figure}

The example galaxies used in Figure~\ref{Fig:chemical} have different sizes, masses, and metallicity ranges. Looking at the metallicities at which they overlap (i.e. [Fe/H]~$\lesssim-2.0$), a few elements present different [X/Fe] distributions (e.g. C, N, and Ba).
The same variation could exist in all galaxies of our simulations.
In this subsection, we test whether or not the early chemical evolution in our simulated galaxies is a universal process, which  leads to the same [X/Fe] distributions, and is sampled differently in each system. To answer this question, we create a pool containing very metal-poor stars (VMPs, [Fe/H]~$<-2.0$) from each galaxy.
Although we said metal-poor stars are not necessarily old stars in Section~\ref{Sec:agemet}, we still use metallicity as a proxy for ages here, just because stellar ages are difficult to measure in observations.
For this test, we only use the high-resolution run.

Once the VMP pool has been defined, we apply a Kolmogorov–Smirnov (KS) test to assess whether the elemental abundance distributions of the VMP tail of each galaxy are statistically consistent with being randomly drawn from the parent [X/Fe] distributions represented by the VMP pool. This assumption constitutes our null hypothesis. For each elemental abundance ratio [X/Fe], we obtain a distribution of the KS p-values. As a reminder, small p-values, usually $\lesssim0.05$ or $\lesssim0.1$, imply that the null hypothesis is very unlikely. For the 13 elements (including Fe) shown in Figure~\ref{Fig:chemical}, the fraction of galaxies with p-values $< 0.05$  and $<0.10$ is calculated to find how likely the null hypothesis is rejected among the suite.

Figure~\ref{Fig:kolmogorov} shows the fraction of the null hypothesis rejection for each element and for the two thresholds considered (0.05 and 0.10 marked with blue circles and black squares, respectively).
The null hypothesis is largely rejected in all the considered elemental abundance ratios, meaning [X/Fe] distributions are significantly different among galaxies. Mg and Mn are those elements for which the null hypothesis rejection fraction is the smallest, i.e. 54 and 52 percent, respectively, if 0.05 is considered as a threshold. This is probably because these two elements have the lowest elemental abundance dispersions relative to Fe (as shown in Figure~\ref{Fig:chemical}). The overall high values of the null hypothesis rejection fraction, independent of the threshold of the KS test, suggest that the early chemical evolution in these simulated galaxies proceed differently, leading to a variegated ranges of [X/Fe] ratios and different distributions (as shown in Figure~\ref{Fig:chemical}).
We did not find a clear correlation of the p-values against galaxy mass.

\section{Summary}\label{Sec:conclusions}
In this paper, we present the chemo-dynamical properties of satellite dwarf galaxy suites detected in our cosmological hydrodynamical zoom-in simulations of a Milky Way-type galaxy. 
Here we summarise the main properties and our findings: 
\begin{enumerate}
    \item Using HDBSCAN algorithm, we identify 223 and 118 dwarf galaxies 
    with $3\times10^4 \lesssim \rm{M_\star/M_{\odot} }\lesssim 6\times10^{10} $  in the high (Aq-C-4) and medium (Aq-C-5) resolution runs, respectively (see Figure~\ref{Fig:nice} for the example).
    
    \item The spatial 3D distribution of the satellite galaxies follows a cosmic filament across $\sim11-12$ Mpc, in which a Milky Way-like galaxy is at the centre (Figure~\ref{Fig:position}).
    
    \item There is a clear relation between stellar mass, gas mass, and dark matter mass in the simulated galaxies (Figure~\ref{Fig:mass}). Our suites provide galaxies with a large range of gas-to-star mass ratios, from 0 to few $10^2$.
    
    \item Our suites cover a wide range of star formation histories. The time at which half of the stars are formed, $t_{50}$, present a wide ranges of values: those that formed stars only earlier on (shorter $t_{50}$) and those that had a more continuos star formation (longer $t_{50}$).
    We find a complex trend with the systemic metallicity and the stellar masses, in good agreement with the derived values from observed Local Group galaxies (Figure~\ref{Fig:starformation}). The general trend is that more massive galaxies present a longer $t_{50}$.  
    
    \item The mean systemic metallicities, the systemic velocity dispersions, the luminosities, and the half-light radii of the simulated galaxies are in excellent agreement with the values of the observed Local Group galaxies  at $M_\star \gtrsim 10^8$ (see Figure~\ref{Fig:various})
    At the lower mass, as in many other cosmological simulations, the  mean metallicities and sizes of the least massive galaxies are under-predicted and over-predicted respectively, which, we find, is likely caused by astrophysics rather than numerical resolutions. 
    
    \item There is a clear trend between ages and metallicities of member stars (chemical evolution), which also changes as a function of the system stellar mass (Figure~\ref{Fig:agemet}). The least-massive systems does not reach solar metallicities, and more massive systems reach higher metallicities at earlier cosmic times.
    In a given system, the oldest stars are, generally, also the most metal-poor stars. However, we find late-type, metal-poor galaxies, and metal poor stars are not necessarily the old stars. This challenges the fundamental assumption in the Galactic archaeology studies.
    
    \item For the first time in a high-resolution cosmological simulation, stellar particles also posses elemental abundances of all stable elements (from hydrogen to uranium). Our predicted elemental abundances span in a wide range as a function of metallicities (Figure~\ref{Fig:chemical}). The lowest mass systems tend to be dominated by inhomogeneous mixing and are more pristine, while the more massive ones are more homogeneous chemical evolution also showing Type~Ia supernovae enrichment.

    \item At [Fe/H] $<-2$, the elemental abundance dispersions in satellite galaxies compared with the central galaxy (quantified as the MAD ratio) has a clear trend against the systemic stellar masses (Figure~\ref{Fig:madratio}). The low-mass satellites with $\rm{M_\star/M_{\odot} } \lesssim 10^6$ are dominated by inhomogeneous mixing star formation, which causes a large elemental abundance dispersion. As the systemic mass increases, the chemical evolution proceeds more homogeneously, lowering the dispersions. In this regime the Milky Way-like galaxy have higher elemental abundance dispersions than those of dwarf galaxies (MAD ratios $>1$). In the most massive satellites with $\rm{M_\star/M_{\odot} }\gtrsim 10^9$, the elemental abundance dispersions  are similar to those of the simulated Milky Way, as in both cases, the systems are built from  various accreted systems with different chemical evolution. The detailed accretion history will be analysed in our next paper (Sestito~\&~Kobayashi, in prep.).
    
    \item The Kolmogorov-Smirnov test suggests that the chemical evolution in the simulated galaxies is very likely to have proceeded differently, even at the earliest epochs (Figure~\ref{Fig:kolmogorov}). We did not find a clear relationship against systemic mass.
    
\end{enumerate}

Various multi-object spectroscopic (MOS) surveys recently started, i.e. DESI \citep{Cooper23,DESIstellarcat}, the Subaru $\onohiula$ PFS-SSP  \citep{PFS,PFSGA}, and 4MOST \citep[e.g.][]{deJong19,4DWARFS}, which will soon be followed by WEAVE \citep[e.g.][]{WEAVE12,WEAVE24}. All of them will provide spectroscopic information for millions of stars in the Local Group. Our simulations include all stable elements, and our suites of dwarf satellite galaxies, that span a large range of chemo-dynamical properties, will be very useful to interpret the upcoming large datasets, to then explore the formation and evolution of the Local Group, and to constrain the nucleosynthetic channels. This has been already anticipated in \citet{Sestito26b}, in which the simulated galaxy Aq-C-4-sub182 has been used to interpret the metallicity and the velocity distributions of the Sextans dwarf galaxy obtained from DESI DR1 data.

\section*{Acknowledgements}
FS and CK acknowledge funding from the UK Science and Technology Facilities Council through grant ST/Y001443/1. 
This work has made use of the University of Hertfordshire's high-performance computing facility \url{http://uhhpc.herts.ac.uk/}.
This work used the DiRAC Memory Intensive service (Cosma8 / Cosma7 / Cosma6 [*]) at Durham University, managed by the Institute for Computational Cosmology on behalf of the STFC DiRAC HPC Facility (www.dirac.ac.uk). The DiRAC service at Durham was funded by BEIS, UKRI and STFC capital funding, Durham University and STFC operations grants. DiRAC is part of the UKRI Digital Research Infrastructure.

This project makes use of the MaNGA-Pipe3D dataproducts. We thank the IA-UNAM MaNGA team for creating this catalogue, and the Conacyt Project CB-285080 for supporting them.

\section*{Data Availability}
The properties of the simulated galaxies will be available as online material. Simulations will be shared upon reasonable request.

\bibliographystyle{mn2e}
\bibliography{aquarius_dwarf}

\begin{thebibliography}{104}
\expandafter\ifx\csname natexlab\endcsname\relax\def\natexlab#1{#1}\fi

\bibitem[{{Arcones}, {Janka} \& {Scheck}(2007){Arcones}, {Janka}, \&
  {Scheck}}]{Arcones07}
{Arcones} A., {Janka} H.-T., {Scheck} L., 2007, \aap, 467, 1227

\bibitem[{{Bastian} \& {Lardo}(2018)}]{Bastian18}
{Bastian} N., {Lardo} C., 2018, \araa, 56, 83

\bibitem[{{Bliss} {et~al}\mbox{.}(2017){Bliss}, {Arcones}, {Montes}, \&
  {Pereira}}]{Bliss17}
{Bliss} J., {Arcones} A., {Montes} F., {Pereira} J., 2017, Journal of Physics G
  Nuclear Physics, 44, 054003

\bibitem[{{Bonifacio} {et~al}\mbox{.}(2025){Bonifacio}, {Caffau},
  {Fran{\c{c}}ois}, {Martin}, {Ibata}, {Yuan}, {Kordopatis}, {Gonz{\'a}lez
  Hern{\'a}ndez}, {Aguado}, {Thomas}, {Viswanathan}, {Dodd}, {Gran},
  {Starkenburg}, {Lardo}, {Errani}, {Fouesneau}, {Navarro}, {Venn}, \&
  {Malhan}}]{Bonifacio25}
{Bonifacio} P. {et~al.}, 2025, \aap, 695, A113

\bibitem[{{Bromm}(2013)}]{Bromm13}
{Bromm} V., 2013, Reports on Progress in Physics, 76, 112901

\bibitem[{{Buck} {et~al}\mbox{.}(2020){Buck}, {Obreja}, {Macci{\`o}},
  {Minchev}, {Dutton}, \& {Ostriker}}]{Buck20}
{Buck} T., {Obreja} A., {Macci{\`o}} A.~V., {Minchev} I., {Dutton} A.~A.,
  {Ostriker} J.~P., 2020, \mnras, 491, 3461

\bibitem[{{Buder} {et~al}\mbox{.}(2025){Buder}, {Buck}, {Sk{\'u}lad{\'o}ttir},
  {Ness}, {McKenzie}, \& {Monty}}]{Buder25}
{Buder} S., {Buck} T., {Sk{\'u}lad{\'o}ttir} {\'A}., {Ness} M., {McKenzie} M.,
  {Monty} S., 2025, arXiv e-prints, arXiv:2510.11284

\bibitem[{{Bullock} \& {Boylan-Kolchin}(2017)}]{Bullock17}
{Bullock} J.~S., {Boylan-Kolchin} M., 2017, \araa, 55, 343

\bibitem[{{Carlberg} {et~al}\mbox{.}(2025){Carlberg}, {Ibata}, {Martin},
  {Starkenburg}, {Aguado}, {Malhan}, {Venn}, \& {Yuan}}]{Carlberg25}
{Carlberg} R.~G., {Ibata} R., {Martin} N.~F., {Starkenburg} E., {Aguado} D.~S.,
  {Malhan} K., {Venn} K., {Yuan} Z., 2025, \apj, 988, 96

\bibitem[{{Chiba} {et~al}\mbox{.}(2026){Chiba}, {Wyse}, {Kirby}, {Cohen},
  {Dobos}, {Gerasimov}, {Ishigaki}, {Hayashi}, {Filion}, {Arnaboldi},
  {Bhattacharya}, {Hirai}, {Kobayashi}, {Komiyama}, {Kuzma}, {Ogami},
  {Chies-Santos}, {Klock-Miranda}, {Sestito}, {Budav{\'a}ri}, {Cooper}, {Ding},
  {Escala}, {Ferreira}, {Gerhard}, {Henderson}, {Hong}, {Horigome}, {Ikeda},
  {Ishikawa}, {Kirihara}, {Li}, {Martin}, {Miyazaki Sakurako Okamoto},
  {Pattnaik}, {Sato}, {Suzuki}, {Szalay}, {Wardana}, {Wei}, {Wu}, {Wu}, {Xu},
  {Ye}, {Miki}, {Zhang}, {Zhao}, {Zhao}, \& {Zhao}}]{PFSGA}
{Chiba} M. {et~al.}, 2026, arXiv e-prints, arXiv:2604.09875

\bibitem[{{Cooper} {et~al}\mbox{.}(2023){Cooper}, {Koposov}, {Allende Prieto},
  {Manser}, {Kizhuprakkat}, {Myers}, {Dey}, {G{\"a}nsicke}, {Li}, {Rockosi},
  {Valluri}, {Najita}, {Deason}, {Raichoor}, {Wang}, {Ting}, {Kim}, {Carrillo},
  {Wang}, {Beraldo e Silva}, {Han}, {Ding}, {S{\'a}nchez-Conde}, {Aguilar},
  {Ahlen}, {Bailey}, {Belokurov}, {Brooks}, {Cunha}, {Dawson}, {de la Macorra},
  {Doel}, {Eisenstein}, {Fagrelius}, {Fanning}, {Font-Ribera}, {Forero-Romero},
  {Gazta{\~n}aga}, {Gontcho a Gontcho}, {Guy}, {Honscheid}, {Kehoe}, {Kisner},
  {Kremin}, {Landriau}, {Levi}, {Martini}, {Meisner}, {Miquel}, {Moustakas},
  {Nie}, {Palanque-Delabrouille}, {Percival}, {Poppett}, {Prada}, {Rehemtulla},
  {Schlafly}, {Schlegel}, {Schubnell}, {Sharples}, {Tarl{\'e}}, {Wechsler},
  {Weinberg}, {Zhou}, \& {Zou}}]{Cooper23}
{Cooper} A.~P. {et~al.}, 2023, \apj, 947, 37

\bibitem[{{Dalton} {et~al}\mbox{.}(2012){Dalton}, {Trager}, {Abrams}, {Carter},
  {Bonifacio}, {Aguerri}, {MacIntosh}, {Evans}, {Lewis}, {Navarro}, {Agocs},
  {Dee}, {Rousset}, {Tosh}, {Middleton}, {Pragt}, {Terrett}, {Brock}, {Benn},
  {Verheijen}, {Cano Infantes}, {Bevil}, {Steele}, {Mottram}, {Bates},
  {Gribbin}, {Rey}, {Rodriguez}, {Delgado}, {Guinouard}, {Walton}, {Irwin},
  {Jagourel}, {Stuik}, {Gerlofsma}, {Roelfsma}, {Skillen}, {Ridings},
  {Balcells}, {Daban}, {Gouvret}, {Venema}, \& {Girard}}]{WEAVE12}
{Dalton} G. {et~al.}, 2012, Society of Photo-Optical Instrumentation Engineers
  (SPIE) Conference Series, Vol. 8446, {WEAVE: the next generation wide-field
  spectroscopy facility for the William Herschel Telescope}, p. 84460P

\bibitem[{{de Jong} {et~al}\mbox{.}(2019){de Jong}, {Agertz}, {Berbel}, {Aird},
  {Alexander}, {Amarsi}, {Anders}, {Andrae}, {Ansarinejad}, {Ansorge},
  {Antilogus}, {Anwand -Heerwart}, {Arentsen}, {Arnadottir}, {Asplund},
  {Auger}, {Azais}, {Baade}, {Baker}, {Baker}, {Balbinot}, {Baldry}, {Banerji},
  {Barden}, {Barklem}, {Barth{\'e}l{\'e}my-Mazot}, {Battistini}, {Bauer},
  {Bell}, {Bellido-Tirado}, {Bellstedt}, {Belokurov}, {Bensby}, {Bergemann},
  {Bestenlehner}, {Bielby}, {Bilicki}, {Blake}, {Bland-Hawthorn}, {Boeche},
  {Boland}, {Boller}, {Bongard}, {Bongiorno}, {Bonifacio}, {Boudon}, {Brooks},
  {Brown}, {Brown}, {Br{\"u}ggen}, {Brynnel}, {Brzeski}, {Buchert},
  {Buschkamp}, {Caffau}, {Caillier}, {Carrick}, {Casagrande}, {Case}, {Casey},
  {Cesarini}, {Cescutti}, {Chapuis}, {Chiappini}, {Childress}, {Christlieb},
  {Church}, {Cioni}, {Cluver}, {Colless}, {Collett}, {Comparat}, {Cooper},
  {Couch}, {Courbin}, {Croom}, {Croton}, {Daguis{\'e}}, {Dalton}, {Davies},
  {Davis}, {de Laverny}, {Deason}, {Dionies}, {Disseau}, {Doel}, {D{\"o}scher},
  {Driver}, {Dwelly}, {Eckert}, {Edge}, {Edvardsson}, {Youssoufi}, {Elhaddad},
  {Enke}, {Erfanianfar}, {Farrell}, {Fechner}, {Feiz}, {Feltzing}, {Ferreras},
  {Feuerstein}, {Feuillet}, {Finoguenov}, {Ford}, {Fotopoulou}, {Fouesneau},
  {Frenk}, {Frey}, {Gaessler}, {Geier}, {Fusillo}, {Gerhard}, {Giannantonio},
  {Giannone}, {Gibson}, {Gillingham}, {Gonz{\'a}lez-Fern{\'a}ndez},
  {Gonzalez-Solares}, {Gottloeber}, {Gould}, {Grebel}, {Gueguen}, {Guiglion},
  {Haehnelt}, {Hahn}, {Hansen}, {Hartman}, {Hauptner}, {Hawkins}, {Haynes},
  {Haynes}, {Heiter}, {Helmi}, {Aguayo}, {Hewett}, {Hinton}, {Hobbs}, {Hoenig},
  {Hofman}, {Hook}, {Hopgood}, {Hopkins}, {Hourihane}, {Howes}, {Howlett},
  {Huet}, {Irwin}, {Iwert}, {Jablonka}, {Jahn}, {Jahnke}, {Jarno}, {Jin},
  {Jofre}, {Johl}, {Jones}, {J{\"o}nsson}, {Jordan}, {Karovicova}, {Khalatyan},
  {Kelz}, {Kennicutt}, {King}, {Kitaura}, {Klar}, {Klauser}, {Kneib}, {Koch},
  {Koposov}, {Kordopatis}, {Korn}, {Kosmalski}, {Kotak}, {Kovalev}, {Kreckel},
  {Kripak}, {Krumpe}, {Kuijken}, {Kunder}, {Kushniruk}, {Lam}, {Lamer},
  {Laurent}, {Lawrence}, {Lehmitz}, {Lemasle}, {Lewis}, {Li}, {Lidman}, {Lind},
  {Liske}, {Lizon}, {Loveday}, {Ludwig}, {McDermid}, {Maguire}, {Mainieri},
  {Mali}, {Mandel}, {Mandel}, {Mannering}, {Martell}, {Martinez Delgado},
  {Matijevic}, {McGregor}, {McMahon}, {McMillan}, {Mena}, {Merloni}, {Meyer},
  {Michel}, {Micheva}, {Migniau}, {Minchev}, {Monari}, {Muller}, {Murphy},
  {Muthukrishna}, {Nandra}, {Navarro}, {Ness}, {Nichani}, {Nichol}, {Nicklas},
  {Niederhofer}, {Norberg}, {Obreschkow}, {Oliver}, {Owers}, {Pai},
  {Pankratow}, {Parkinson}, {Paschke}, {Paterson}, {Pecontal}, {Parry},
  {Phillips}, {Pillepich}, {Pinard}, {Pirard}, {Piskunov}, {Plank},
  {Pl{\"u}schke}, {Pons}, {Popesso}, {Power}, {Pragt}, {Pramskiy}, {Pryer},
  {Quattri}, {Queiroz}, {Quirrenbach}, {Rahurkar}, {Raichoor}, {Ramstedt},
  {Rau}, {Recio-Blanco}, {Reiss}, {Renaud}, {Revaz}, {Rhode}, {Richard},
  {Richter}, {Rix}, {Robotham}, {Roelfsema}, {Romaniello}, {Rosario},
  {Rothmaier}, {Roukema}, {Ruchti}, {Rupprecht}, {Rybizki}, {Ryde}, {Saar},
  {Sadler}, {Sahl{\'e}n}, {Salvato}, {Sassolas}, {Saunders}, {Saviauk},
  {Sbordone}, {Schmidt}, {Schnurr}, {Scholz}, {Schwope}, {Seifert}, {Shanks},
  {Sheinis}, {Sivov}, {Sk{\'u}lad{\'o}ttir}, {Smartt}, {Smedley}, {Smith},
  {Smith}, {Sorce}, {Spitler}, {Starkenburg}, {Steinmetz}, {Stilz}, {Storm},
  {Sullivan}, {Sutherland}, {Swann}, {Tamone}, {Taylor}, {Teillon}, {Tempel},
  {ter Horst}, {Thi}, {Tolstoy}, {Trager}, {Traven}, {Tremblay}, {Tresse},
  {Valentini}, {van de Weygaert}, {van den Ancker}, {Veljanoski}, {Venkatesan},
  {Wagner}, {Wagner}, {Walcher}, {Waller}, {Walton}, {Wang}, {Winkler},
  {Wisotzki}, {Worley}, {Worseck}, {Xiang}, {Xu}, {Yong}, {Zhao}, {Zheng},
  {Zscheyge}, \& {Zucker}}]{deJong19}
{de Jong} R.~S. {et~al.}, 2019, The Messenger, 175, 3

\bibitem[{{Deason}, {Wetzel} \& {Garrison-Kimmel}(2014){Deason}, {Wetzel}, \&
  {Garrison-Kimmel}}]{Deason14}
{Deason} A., {Wetzel} A., {Garrison-Kimmel} S., 2014, \apj, 794, 115

\bibitem[{{El-Badry} {et~al}\mbox{.}(2018{\natexlab{a}}){El-Badry},
  {Bland-Hawthorn}, {Wetzel}, {Quataert}, {Weisz}, {Boylan-Kolchin}, {Hopkins},
  {Faucher-Gigu{\`e}re}, {Kere{\v s}}, \& {Garrison-Kimmel}}]{ElBadry18}
{El-Badry} K. {et~al.}, 2018{\natexlab{a}}, \mnras, 480, 652

\bibitem[{{El-Badry} {et~al}\mbox{.}(2018{\natexlab{b}}){El-Badry}, {Bradford},
  {Quataert}, {Geha}, {Boylan-Kolchin}, {Weisz}, {Wetzel}, {Hopkins}, {Chan},
  {Fitts}, {Kere{\v{s}}}, \& {Faucher-Gigu{\`e}re}}]{ElBadry18b}
---, 2018{\natexlab{b}}, \mnras, 477, 1536

\bibitem[{{Eldridge} {et~al}\mbox{.}(2017){Eldridge}, {Stanway}, {Xiao},
  {McClelland}, {Taylor}, {Ng}, {Greis}, \& {Bray}}]{Eldridge17}
{Eldridge} J.~J., {Stanway} E.~R., {Xiao} L., {McClelland} L.~A.~S., {Taylor}
  G., {Ng} M., {Greis} S.~M.~L., {Bray} J.~C., 2017, \pasa, 34, e058

\bibitem[{{Errani} {et~al}\mbox{.}(2022){Errani}, {Navarro}, {Ibata}, {Martin},
  {Yuan}, {Aguado}, {Bonifacio}, {Caffau}, {Gonz{\'a}lez Hern{\'a}ndez},
  {Malhan}, {S{\'a}nchez-Janssen}, {Sestito}, {Starkenburg}, {Thomas}, \&
  {Venn}}]{Errani22b}
{Errani} R. {et~al.}, 2022, \mnras, 514, 3532

\bibitem[{{Fattahi} {et~al}\mbox{.}(2018){Fattahi}, {Navarro}, {Frenk}, {Oman},
  {Sawala}, \& {Schaller}}]{Fattahi18}
{Fattahi} A., {Navarro} J.~F., {Frenk} C.~S., {Oman} K.~A., {Sawala} T.,
  {Schaller} M., 2018, \mnras, 476, 3816

\bibitem[{{Feltzing} {et~al}\mbox{.}(2009){Feltzing}, {Eriksson}, {Kleyna}, \&
  {Wilkinson}}]{Feltzing09}
{Feltzing} S., {Eriksson} K., {Kleyna} J., {Wilkinson} M.~I., 2009, \aap, 508,
  L1

\bibitem[{{Frebel} \& {Norris}(2015)}]{Frebel15}
{Frebel} A., {Norris} J.~E., 2015, \araa, 53, 631

\bibitem[{{Frebel} {et~al}\mbox{.}(2016){Frebel}, {Norris}, {Gilmore}, \&
  {Wyse}}]{Frebel16}
{Frebel} A., {Norris} J.~E., {Gilmore} G., {Wyse} R. F.~G., 2016, \apj, 826,
  110

\bibitem[{{Gilmore} {et~al}\mbox{.}(2013){Gilmore}, {Norris}, {Monaco}, {Yong},
  {Wyse}, \& {Geisler}}]{Gilmore13}
{Gilmore} G., {Norris} J.~E., {Monaco} L., {Yong} D., {Wyse} R. F.~G.,
  {Geisler} D., 2013, \apj, 763, 61

\bibitem[{{Governato} {et~al}\mbox{.}(2010){Governato}, {Brook}, {Mayer},
  {Brooks}, {Rhee}, {Wadsley}, {Jonsson}, {Willman}, {Stinson}, {Quinn}, \&
  {Madau}}]{Governato10}
{Governato} F. {et~al.}, 2010, \nat, 463, 203

\bibitem[{{Grebel}, {Gallagher} \& {Harbeck}(2003){Grebel}, {Gallagher}, \&
  {Harbeck}}]{Grebel03}
{Grebel} E.~K., {Gallagher}, John~S. I., {Harbeck} D., 2003, \aj, 125, 1926

\bibitem[{{Haynes} \& {Kobayashi}(2019)}]{Haynes19}
{Haynes} C.~J., {Kobayashi} C., 2019, \mnras, 483, 5123

\bibitem[{{Heiger} {et~al}\mbox{.}(2026){Heiger}, {Ji}, {Li}, {Simon},
  {Limberg}, {Carballo-Bello}, {Cerny}, {Chaturvedi}, {Chiti}, {Choi},
  {Crnojevi{\'c}}, {Mart{\'\i}nez-V{\'a}zquez}, {Medina}, {Mutlu-Pakdil},
  {Navabi}, {No{\"e}l}, {Pace}, {Placco}, {Riley}, {Sakowska}, {Stringfellow},
  \& {Delve Collaboration}}]{Heiger26}
{Heiger} M.~E. {et~al.}, 2026, \apj, 1002, 182

\bibitem[{{Higgs} {et~al}\mbox{.}(2021){Higgs}, {McConnachie}, {Annau},
  {Irwin}, {Battaglia}, {C{\^o}t{\'e}}, {Lewis}, \& {Venn}}]{Higgs21}
{Higgs} C.~R., {McConnachie} A.~W., {Annau} N., {Irwin} M., {Battaglia} G.,
  {C{\^o}t{\'e}} P., {Lewis} G.~F., {Venn} K., 2021, \mnras, 503, 176

\bibitem[{{Hirai} {et~al}\mbox{.}(2024){Hirai}, {Kirby}, {Chiba}, {Hayashi},
  {Anguiano}, {Saitoh}, {Ishigaki}, \& {Beers}}]{Hirai24}
{Hirai} Y., {Kirby} E.~N., {Chiba} M., {Hayashi} K., {Anguiano} B., {Saitoh}
  T.~R., {Ishigaki} M.~N., {Beers} T.~C., 2024, \apj, 970, 105

\bibitem[{{Hopkins} {et~al}\mbox{.}(2014){Hopkins}, {Kere{\v{s}}},
  {O{\~n}orbe}, {Faucher-Gigu{\`e}re}, {Quataert}, {Murray}, \&
  {Bullock}}]{Hopkins14}
{Hopkins} P.~F., {Kere{\v{s}}} D., {O{\~n}orbe} J., {Faucher-Gigu{\`e}re}
  C.-A., {Quataert} E., {Murray} N., {Bullock} J.~S., 2014, \mnras, 445, 581

\bibitem[{{Ishigaki} {et~al}\mbox{.}(2014){Ishigaki}, {Aoki}, {Arimoto}, \&
  {Okamoto}}]{Ishigaki14}
{Ishigaki} M.~N., {Aoki} W., {Arimoto} N., {Okamoto} S., 2014, \aap, 562, A146

\bibitem[{{Izutani}, {Umeda} \& {Tominaga}(2009){Izutani}, {Umeda}, \&
  {Tominaga}}]{Izutani09}
{Izutani} N., {Umeda} H., {Tominaga} N., 2009, \apj, 692, 1517

\bibitem[{{Jensen} {et~al}\mbox{.}(2024){Jensen}, {Hayes}, {Sestito},
  {McConnachie}, {Waller}, {Smith}, {Navarro}, \& {Venn}}]{Jensen24}
{Jensen} J., {Hayes} C.~R., {Sestito} F., {McConnachie} A.~W., {Waller} F.,
  {Smith} S. E.~T., {Navarro} J., {Venn} K.~A., 2024, \mnras, 527, 4209

\bibitem[{{Ji} {et~al}\mbox{.}(2019){Ji}, {Simon}, {Frebel}, {Venn}, \&
  {Hansen}}]{Ji19}
{Ji} A.~P., {Simon} J.~D., {Frebel} A., {Venn} K.~A., {Hansen} T.~T., 2019,
  \apj, 870, 83

\bibitem[{{Jin} {et~al}\mbox{.}(2024){Jin}, {Trager}, {Dalton}, {Aguerri},
  {Drew}, {Falc{\'o}n-Barroso}, {G{\"a}nsicke}, {Hill}, {Iovino}, {Pieri}, \&
  et~al.}]{WEAVE24}
{Jin} S. {et~al.}, 2024, \mnras, 530, 2688

\bibitem[{{Kazantzidis} {et~al}\mbox{.}(2011){Kazantzidis}, {{\L}okas},
  {Callegari}, {Mayer}, \& {Moustakas}}]{Kazantzidis11}
{Kazantzidis} S., {{\L}okas} E.~L., {Callegari} S., {Mayer} L., {Moustakas}
  L.~A., 2011, \apj, 726, 98

\bibitem[{{Kirby} {et~al}\mbox{.}(2013){Kirby}, {Cohen}, {Guhathakurta},
  {Cheng}, {Bullock}, \& {Gallazzi}}]{Kirby13}
{Kirby} E.~N., {Cohen} J.~G., {Guhathakurta} P., {Cheng} L., {Bullock} J.~S.,
  {Gallazzi} A., 2013, \apj, 779, 102

\bibitem[{{Kirby} {et~al}\mbox{.}(2011){Kirby}, {Lanfranchi}, {Simon}, {Cohen},
  \& {Guhathakurta}}]{Kirby11}
{Kirby} E.~N., {Lanfranchi} G.~A., {Simon} J.~D., {Cohen} J.~G., {Guhathakurta}
  P., 2011, \apj, 727, 78

\bibitem[{{Kobayashi}(2004)}]{Kobayashi04}
{Kobayashi} C., 2004, \mnras, 347, 740

\bibitem[{{Kobayashi}(2025)}]{Kobayashi25}
---, 2025, arXiv e-prints, arXiv:2506.20436

\bibitem[{{Kobayashi}, {Karakas} \& {Lugaro}(2020){Kobayashi}, {Karakas}, \&
  {Lugaro}}]{Kobayashi20}
{Kobayashi} C., {Karakas} A.~I., {Lugaro} M., 2020, \apj, 900, 179

\bibitem[{{Kobayashi}, {Leung} \& {Nomoto}(2020){Kobayashi}, {Leung}, \&
  {Nomoto}}]{Kobayashi20b}
{Kobayashi} C., {Leung} S.-C., {Nomoto} K., 2020, \apj, 895, 138

\bibitem[{{Kobayashi} {et~al}\mbox{.}(2023){Kobayashi}, {Mandel}, {Belczynski},
  {Goriely}, {Janka}, {Just}, {Ruiter}, {Vanbeveren}, {Kruckow}, {Briel},
  {Eldridge}, \& {Stanway}}]{Kobayashi23}
{Kobayashi} C. {et~al.}, 2023, \apjl, 943, L12

\bibitem[{{Kobayashi} \& {Nakasato}(2011)}]{Kobayashi11b}
{Kobayashi} C., {Nakasato} N., 2011, \apj, 729, 16

\bibitem[{{Kobayashi} \& {Nomoto}(2009)}]{Kobayashi09}
{Kobayashi} C., {Nomoto} K., 2009, \apj, 707, 1466

\bibitem[{{Kobayashi}, {Springel} \& {White}(2007){Kobayashi}, {Springel}, \&
  {White}}]{Kobayashi07}
{Kobayashi} C., {Springel} V., {White} S. D.~M., 2007, \mnras, 376, 1465

\bibitem[{{Kobayashi} {et~al}\mbox{.}(2006){Kobayashi}, {Umeda}, {Nomoto},
  {Tominaga}, \& {Ohkubo}}]{Kobayashi06}
{Kobayashi} C., {Umeda} H., {Nomoto} K., {Tominaga} N., {Ohkubo} T., 2006,
  \apj, 653, 1145

\bibitem[{{Komatsu} {et~al}\mbox{.}(2009){Komatsu}, {Dunkley}, {Nolta},
  {Bennett}, {Gold}, {Hinshaw}, {Jarosik}, {Larson}, {Limon}, {Page},
  {Spergel}, {Halpern}, {Hill}, {Kogut}, {Meyer}, {Tucker}, {Weiland},
  {Wollack}, \& {Wright}}]{Komatsu09}
{Komatsu} E. {et~al.}, 2009, \apjs, 180, 330

\bibitem[{{Koposov} {et~al}\mbox{.}(2026){Koposov}, {Li}, {Prieto}, {Medina},
  {Sandford}, {Aguado}, {Silva}, {Bystr{\"o}m}, {Cooper}, {Dey}, {Frenk},
  {Kizhuprakkat}, {Li}, {Najita}, {Riley}, {Silva}, {Thomas}, {Valluri},
  {Aguilar}, {Ahlen}, {Bianchi}, {Brooks}, {Claybaugh}, {Cole}, {Cuceu}, {de la
  Macorra}, {Della Costa}, {Dey}, {Doel}, {Font-Ribera}, {Edelstein},
  {Forero-Romero}, {Gazta{\~n}aga}, {Gontcho}, {Gutierrez}, {Guy}, {Honscheid},
  {Jimenez}, {Kehoe}, {Kirkby}, {Kisner}, {Kremin}, {Lahav}, {Landriau},
  {Guillou}, {Leauthaud}, {Levi}, {Manera}, {Meisner}, {Miquel}, {Moustakas},
  {Nadathur}, {Palanque-Delabrouille}, {Percival}, {Prada},
  {P{\'e}rez-R{\`a}fols}, {Rossi}, {Sanchez}, {Schlafly}, {Schlegel}, {Seo},
  {Sharples}, {Silber}, {Sprayberry}, {Tarl'e}, {Weaver}, {Zhou}, \&
  {Zou}}]{DESIstellarcat}
{Koposov} S. {et~al.}, 2026, The Open Journal of Astrophysics, 9, 55260

\bibitem[{{Kroupa}(2002)}]{Kroupa02}
{Kroupa} P., 2002, Science, 295, 82

\bibitem[{{Lacerda} {et~al}\mbox{.}(2022){Lacerda}, {S{\'a}nchez},
  {Mej{\'\i}a-Narv{\'a}ez}, {Camps-Fari{\~n}a}, {Espinosa-Ponce},
  {Barrera-Ballesteros}, {Ibarra-Medel}, \& {Lugo-Aranda}}]{Lacerda22}
{Lacerda} E. A.~D., {S{\'a}nchez} S.~F., {Mej{\'\i}a-Narv{\'a}ez} A.,
  {Camps-Fari{\~n}a} A., {Espinosa-Ponce} C., {Barrera-Ballesteros} J.~K.,
  {Ibarra-Medel} H., {Lugo-Aranda} A.~Z., 2022, \na, 97, 101895

\bibitem[{{Limongi} \& {Chieffi}(2018)}]{Limongi18}
{Limongi} M., {Chieffi} A., 2018, \apjs, 237, 13

\bibitem[{{Macci{\`o}} {et~al}\mbox{.}(2017){Macci{\`o}}, {Frings}, {Buck},
  {Penzo}, {Dutton}, {Blank}, \& {Obreja}}]{Maccio17}
{Macci{\`o}} A.~V., {Frings} J., {Buck} T., {Penzo} C., {Dutton} A.~A., {Blank}
  M., {Obreja} A., 2017, \mnras, 472, 2356

\bibitem[{{Martin} {et~al}\mbox{.}(2022){Martin}, {Venn}, {Aguado},
  {Starkenburg}, {Gonz{\'a}lez Hern{\'a}ndez}, {Ibata}, {Bonifacio}, {Caffau},
  {Sestito}, {Arentsen}, {Allende Prieto}, {Carlberg}, {Fabbro}, {Fouesneau},
  {Hill}, {Jablonka}, {Kordopatis}, {Lardo}, {Malhan}, {Mashonkina},
  {McConnachie}, {Navarro}, {S{\'a}nchez-Janssen}, {Thomas}, {Yuan}, \&
  {Mucciarelli}}]{Martin22}
{Martin} N.~F. {et~al.}, 2022, \nat, 601, 45

\bibitem[{{Mashonkina} {et~al}\mbox{.}(2017){Mashonkina}, {Jablonka},
  {Sitnova}, {Pakhomov}, \& {North}}]{Mashonkina17b}
{Mashonkina} L., {Jablonka} P., {Sitnova} T., {Pakhomov} Y., {North} P., 2017,
  \aap, 608, A89

\bibitem[{{Mateo}(1998)}]{Mateo98}
{Mateo} M.~L., 1998, \araa, 36, 435

\bibitem[{{McConnachie}(2012)}]{Mcconnachie12}
{McConnachie} A.~W., 2012, \aj, 144, 4

\bibitem[{McInnes, Healy \& Astels(2017)McInnes, Healy, \&
  Astels}]{McInnes2017}
McInnes L., Healy J., Astels S., 2017, The Journal of Open Source Software, 2

\bibitem[{{McWilliam}, {Wallerstein} \& {Mottini}(2013){McWilliam},
  {Wallerstein}, \& {Mottini}}]{McWilliam13}
{McWilliam} A., {Wallerstein} G., {Mottini} M., 2013, \apj, 778, 149

\bibitem[{{Nishimura}, {Takiwaki} \& {Thielemann}(2015){Nishimura}, {Takiwaki},
  \& {Thielemann}}]{Nishimura15}
{Nishimura} N., {Takiwaki} T., {Thielemann} F.-K., 2015, \apj, 810, 109

\bibitem[{{Norris} {et~al}\mbox{.}(2010){Norris}, {Wyse}, {Gilmore}, {Yong},
  {Frebel}, {Wilkinson}, {Belokurov}, \& {Zucker}}]{Norris10}
{Norris} J.~E., {Wyse} R. F.~G., {Gilmore} G., {Yong} D., {Frebel} A.,
  {Wilkinson} M.~I., {Belokurov} V., {Zucker} D.~B., 2010, \apj, 723, 1632

\bibitem[{{Pignatari} {et~al}\mbox{.}(2008){Pignatari}, {Gallino}, {Meynet},
  {Hirschi}, {Herwig}, \& {Wiescher}}]{Pignatari08}
{Pignatari} M., {Gallino} R., {Meynet} G., {Hirschi} R., {Herwig} F.,
  {Wiescher} M., 2008, \apjl, 687, L95

\bibitem[{{Raiteri} {et~al}\mbox{.}(1991){Raiteri}, {Busso}, {Gallino},
  {Picchio}, \& {Pulone}}]{Raiteri91}
{Raiteri} C.~M., {Busso} M., {Gallino} R., {Picchio} G., {Pulone} L., 1991,
  \apj, 367, 228

\bibitem[{{Revaz} \& {Jablonka}(2018)}]{Revaz18}
{Revaz} Y., {Jablonka} P., 2018, \aap, 616, A96

\bibitem[{{Rey} {et~al}\mbox{.}(2026){Rey}, {Katz}, {Cadiou}, {Sanati},
  {Agertz}, {Blaizot}, {Cameron}, {Choustikov}, {Devriendt}, {Hauk}, \&
  et~al.}]{Rey26}
{Rey} M. {et~al.}, 2026, The Open Journal of Astrophysics, 9, 69605

\bibitem[{{Sanati} {et~al}\mbox{.}(2023){Sanati}, {Jeanquartier}, {Revaz}, \&
  {Jablonka}}]{Sanati23}
{Sanati} M., {Jeanquartier} F., {Revaz} Y., {Jablonka} P., 2023, \aap, 669, A94

\bibitem[{{S{\'a}nchez} {et~al}\mbox{.}(2022){S{\'a}nchez},
  {Barrera-Ballesteros}, {Lacerda}, {Mej{\'\i}a-Narvaez}, {Camps-Fari{\~n}a},
  {Bruzual}, {Espinosa-Ponce}, {Rodr{\'\i}guez-Puebla}, {Calette},
  {Ibarra-Medel}, {Avila-Reese}, {Hernandez-Toledo}, {Bershady}, {Cano-Diaz},
  \& {Munguia-Cordova}}]{Sanchez22}
{S{\'a}nchez} S.~F. {et~al.}, 2022, \apjs, 262, 36

\bibitem[{{Santos-Santos}, {Frenk} \& {Navarro}(2026){Santos-Santos}, {Frenk},
  \& {Navarro}}]{SantosSantos26}
{Santos-Santos} I. M.~E., {Frenk} C.~S., {Navarro} J.~F., 2026, arXiv e-prints,
  arXiv:2604.09539

\bibitem[{{Scannapieco} {et~al}\mbox{.}(2012){Scannapieco}, {Wadepuhl},
  {Parry}, {Navarro}, {Jenkins}, {Springel}, {Teyssier}, {Carlson}, {Couchman},
  {Crain}, {Dalla Vecchia}, {Frenk}, {Kobayashi}, {Monaco}, {Murante},
  {Okamoto}, {Quinn}, {Schaye}, {Stinson}, {Theuns}, {Wadsley}, {White}, \&
  {Woods}}]{Scannapieco12}
{Scannapieco} C. {et~al.}, 2012, \mnras, 423, 1726

\bibitem[{{Sestito} {et~al}\mbox{.}(2024{\natexlab{a}}){Sestito},
  {Ardern-Arentsen}, {Vitali}, {Montelius}, {Lucchesi}, {Venn}, {Martin},
  {Navarro}, \& {Starkenburg}}]{Sestito24Sgr2}
{Sestito} F. {et~al.}, 2024{\natexlab{a}}, \aap, 690, A333

\bibitem[{{Sestito} {et~al}\mbox{.}(2021){Sestito}, {Buck}, {Starkenburg},
  {Martin}, {Navarro}, {Venn}, {Obreja}, {Jablonka}, \&
  {Macci{\`o}}}]{Sestito21}
---, 2021, \mnras, 500, 3750

\bibitem[{{Sestito} {et~al}\mbox{.}(2026){Sestito}, {Fern{\'a}ndez-Alvar},
  {Brooks}, {Olson}, {Carigi}, {Jofr{\'e}}, {Silva}, {Eldridge}, {Vitali},
  {Venn}, {Hill}, {Ardern-Arentsen}, {Kordopatis}, {Martin}, {Navarro},
  {Starkenburg}, {Tissera}, {Jablonka}, {Lardo}, {Lucchesi}, {Buck}, \&
  {Amayo}}]{Sestito26}
---, 2026, \mnras, 548, stag563

\bibitem[{{Sestito} \& {Kobayashi}(2026)}]{Sestito26b}
{Sestito} F., {Kobayashi} C., 2026, \mnras, 549, stag1137

\bibitem[{{Sestito} {et~al}\mbox{.}(2023{\natexlab{a}}){Sestito}, {Roediger},
  {Navarro}, {Jensen}, {Venn}, {Smith}, {Hayes}, \&
  {McConnachie}}]{Sestito23scl}
{Sestito} F., {Roediger} J., {Navarro} J.~F., {Jensen} J., {Venn} K.~A.,
  {Smith} S. E.~T., {Hayes} C., {McConnachie} A.~W., 2023{\natexlab{a}}, \mnras

\bibitem[{{Sestito} {et~al}\mbox{.}(2024{\natexlab{b}}){Sestito}, {Vitali},
  {Jofre}, {Venn}, {Aguado}, {Aguilera-G{\'o}mez}, {Ardern-Arentsen}, {de Brito
  Silva}, {Carlberg}, {Eldridge}, {Gran}, {Hill}, {Jablonka}, {Kordopatis},
  {Martin}, {Matsuno}, {Rusterucci}, {Starkenburg}, \&
  {Viswanathan}}]{Sestito24Sgr}
{Sestito} F. {et~al.}, 2024{\natexlab{b}}, \aap, 689, A201

\bibitem[{{Sestito} {et~al}\mbox{.}(2023{\natexlab{b}}){Sestito}, {Zaremba},
  {Venn}, {D'Aoust}, {Hayes}, {Jensen}, {Navarro}, {Jablonka},
  {Fern{\'a}ndez-Alvar}, {Glover}, {McConnachie}, \&
  {Chen{\'e}}}]{Sestito23Umi}
---, 2023{\natexlab{b}}, \mnras, 525, 2875

\bibitem[{{Simon}(2019)}]{Simon19}
{Simon} J.~D., 2019, \araa, 57, 375

\bibitem[{{Sk{\'u}lad{\'o}ttir} {et~al}\mbox{.}(2019){Sk{\'u}lad{\'o}ttir},
  {Hansen}, {Salvadori}, \& {Choplin}}]{Skuladottir19}
{Sk{\'u}lad{\'o}ttir} {\'A}., {Hansen} C.~J., {Salvadori} S., {Choplin} A.,
  2019, \aap, 631, A171

\bibitem[{{Sk{\'u}lad{\'o}ttir} {et~al}\mbox{.}(2023){Sk{\'u}lad{\'o}ttir},
  {Puls}, {Amarsi}, {Battaglia}, {Buder}, {Campbell}, {Cardona-Barrero},
  {Christlieb}, {Feuillet}, {Gelli}, {Hansen}, {Hill}, {Ibata}, {Jablonka},
  {Kacharov}, {Karakas}, {Koch-Hansen}, {Lind}, {Lombardo}, {Lucchesi},
  {Lugaro}, {Martin}, {Massari}, {Nordlander}, {Reichert}, {Rossi}, {Ruiter},
  {Salvadori}, {Seitenzahl}, {Tolstoy}, {Xylakis-Dornbusch}, \&
  {Youakim}}]{4DWARFS}
{Sk{\'u}lad{\'o}ttir} {\'A}. {et~al.}, 2023, The Messenger, 190, 19

\bibitem[{{Sk{\'u}lad{\'o}ttir} {et~al}\mbox{.}(2021){Sk{\'u}lad{\'o}ttir},
  {Salvadori}, {Amarsi}, {Tolstoy}, {Irwin}, {Hill}, {Jablonka}, {Battaglia},
  {Starkenburg}, {Massari}, {Helmi}, \& {Posti}}]{Skuladottir21}
---, 2021, \apjl, 915, L30

\bibitem[{{Springel}(2005)}]{Springel05}
{Springel} V., 2005, \mnras, 364, 1105

\bibitem[{{Springel}, {Frenk} \& {White}(2006){Springel}, {Frenk}, \&
  {White}}]{Springel06}
{Springel} V., {Frenk} C.~S., {White} S. D.~M., 2006, \nat, 440, 1137

\bibitem[{{Springel} {et~al}\mbox{.}(2008){Springel}, {Wang}, {Vogelsberger},
  {Ludlow}, {Jenkins}, {Helmi}, {Navarro}, {Frenk}, \& {White}}]{Springel08}
{Springel} V. {et~al.}, 2008, \mnras, 391, 1685

\bibitem[{{Stanway} \& {Eldridge}(2018)}]{Stanway18}
{Stanway} E.~R., {Eldridge} J.~J., 2018, \mnras, 479, 75

\bibitem[{{Starkenburg} {et~al}\mbox{.}(2017){Starkenburg}, {Oman}, {Navarro},
  {Crain}, {Fattahi}, {Frenk}, {Sawala}, \& {Schaye}}]{Starkenburg17a}
{Starkenburg} E., {Oman} K.~A., {Navarro} J.~F., {Crain} R.~A., {Fattahi} A.,
  {Frenk} C.~S., {Sawala} T., {Schaye} J., 2017, \mnras, 465, 2212

\bibitem[{{Stevance}, {Eldridge} \& {Stanway}(2020){Stevance}, {Eldridge}, \&
  {Stanway}}]{Stevance20}
{Stevance} H., {Eldridge} J., {Stanway} E., 2020, The Journal of Open Source
  Software, 5, 1987

\bibitem[{{Taibi} {et~al}\mbox{.}(2022){Taibi}, {Battaglia}, {Leaman},
  {Brooks}, {Riggs}, {Munshi}, {Revaz}, \& {Jablonka}}]{Taibi22}
{Taibi} S., {Battaglia} G., {Leaman} R., {Brooks} A., {Riggs} C., {Munshi} F.,
  {Revaz} Y., {Jablonka} P., 2022, \aap, 665, A92

\bibitem[{{Tamura} {et~al}\mbox{.}(2016){Tamura}, {Takato}, {Shimono},
  {Moritani}, {Yabe}, {Ishizuka}, {Ueda}, {Kamata}, {Aghazarian}, {Arnouts},
  {Barban}, {Barkhouser}, {Borges}, {Braun}, {Carr}, {Chabaud}, {Chang},
  {Chen}, {Chiba}, {Chou}, {Chu}, {Cohen}, {de Almeida}, {de Oliveira}, {de
  Oliveira}, {Dekany}, {Dohlen}, {dos Santos}, {dos Santos}, {Ellis},
  {Fabricius}, {Ferrand}, {Ferreira}, {Golebiowski}, {Greene}, {Gross}, {Gunn},
  {Hammond}, {Harding}, {Hart}, {Heckman}, {Hirata}, {Ho}, {Hope}, {Hovland},
  {Hsu}, {Hu}, {Huang}, {Jaquet}, {Jing}, {Karr}, {Kimura}, {King}, {Komatsu},
  {Le Brun}, {Le F{\`e}vre}, {Le Fur}, {Le Mignant}, {Ling}, {Loomis},
  {Lupton}, {Madec}, {Mao}, {Marrara}, {Mendes de Oliveira}, {Minowa},
  {Morantz}, {Murayama}, {Murray}, {Ohyama}, {Orndorff}, {Pascal}, {Pereira},
  {Reiley}, {Reinecke}, {Ritter}, {Roberts}, {Schwochert}, {Seiffert}, {Smee},
  {Sodre}, {Spergel}, {Steinkraus}, {Strauss}, {Surace}, {Suto}, {Suzuki},
  {Swinbank}, {Tait}, {Takada}, {Tamura}, {Tanaka}, {Tresse}, {Verducci},
  {Vibert}, {Vidal}, {Wang}, {Wen}, {Yan}, \& {Yasuda}}]{PFS}
{Tamura} N. {et~al.}, 2016, in Society of Photo-Optical Instrumentation
  Engineers (SPIE) Conference Series, Vol. 9908, Ground-based and Airborne
  Instrumentation for Astronomy VI, {Evans} C.~J., {Simard} L., {Takami} H.,
  eds., p. 99081M

\bibitem[{{Tolstoy}, {Hill} \& {Tosi}(2009){Tolstoy}, {Hill}, \&
  {Tosi}}]{Tolstoy09}
{Tolstoy} E., {Hill} V., {Tosi} M., 2009, \araa, 47, 371

\bibitem[{{Venn} {et~al}\mbox{.}(2012){Venn}, {Shetrone}, {Irwin}, {Hill},
  {Jablonka}, {Tolstoy}, {Lemasle}, {Divell}, {Starkenburg}, {Letarte},
  {Baldner}, {Battaglia}, {Helmi}, {Kaufer}, \& {Primas}}]{Venn12}
{Venn} K.~A. {et~al.}, 2012, \apj, 751, 102

\bibitem[{{Venn} {et~al}\mbox{.}(2026){Venn}, {Yuan}, {Martin}, {Dovgal},
  {Zaremba}, {Starkenburg}, {Gran}, {Hayes}, {Hill}, {Kobayashi}, {Lardo},
  {McConnachie}, {Matsuno}, {Montelius}, {Placco}, {Sestito},
  {Ardern-Arentsen}, {Battaglia}, {Bonifacio}, {Carlberg}, {Fabbro},
  {Fouesneau}, {Ibata}, {Jablonka}, {Jensen}, {Kordopatis}, {McKenzie},
  {Navarro}, {Pazder}, {Sanchez-Janssen}, {Smith}, {Viswanathan}, {Vitali},
  {Wang}, \& {Wang}}]{Venn26}
---, 2026, arXiv e-prints, arXiv:2603.02445

\bibitem[{{Vincenzo} \& {Kobayashi}(2020)}]{Vincenzo20}
{Vincenzo} F., {Kobayashi} C., 2020, \mnras, 496, 80

\bibitem[{{Waller} {et~al}\mbox{.}(2023){Waller}, {Venn}, {Sestito}, {Jensen},
  {Kielty}, {Borukhovetskaya}, {Hayes}, {McConnachie}, \& {Navarro}}]{Waller23}
{Waller} F. {et~al.}, 2023, \mnras, 519, 1349

\bibitem[{{Wanajo}, {Janka} \& {M{\"u}ller}(2013){Wanajo}, {Janka}, \&
  {M{\"u}ller}}]{Wanajo13}
{Wanajo} S., {Janka} H.-T., {M{\"u}ller} B., 2013, \apjl, 767, L26

\bibitem[{{Wanajo} {et~al}\mbox{.}(2014){Wanajo}, {Sekiguchi}, {Nishimura},
  {Kiuchi}, {Kyutoku}, \& {Shibata}}]{Wanajo14}
{Wanajo} S., {Sekiguchi} Y., {Nishimura} N., {Kiuchi} K., {Kyutoku} K.,
  {Shibata} M., 2014, \apjl, 789, L39

\bibitem[{{Weisz} {et~al}\mbox{.}(2014){Weisz}, {Dolphin}, {Skillman},
  {Holtzman}, {Gilbert}, {Dalcanton}, \& {Williams}}]{Weisz14}
{Weisz} D.~R., {Dolphin} A.~E., {Skillman} E.~D., {Holtzman} J., {Gilbert}
  K.~M., {Dalcanton} J.~J., {Williams} B.~F., 2014, \apj, 789, 147

\bibitem[{{Wetzel} {et~al}\mbox{.}(2016){Wetzel}, {Hopkins}, {Kim},
  {Faucher-Gigu{\`e}re}, {Kere{\v{s}}}, \& {Quataert}}]{Wetzel16}
{Wetzel} A.~R., {Hopkins} P.~F., {Kim} J.-h., {Faucher-Gigu{\`e}re} C.-A.,
  {Kere{\v{s}}} D., {Quataert} E., 2016, \apjl, 827, L23

\bibitem[{{Wheeler} {et~al}\mbox{.}(2019){Wheeler}, {Hopkins}, {Pace},
  {Garrison-Kimmel}, {Boylan-Kolchin}, {Wetzel}, {Bullock}, {Kere{\v{s}}},
  {Faucher-Gigu{\`e}re}, \& {Quataert}}]{Wheeler19}
{Wheeler} C. {et~al.}, 2019, \mnras, 490, 4447

\bibitem[{{White} \& {Rees}(1978)}]{White78}
{White} S.~D.~M., {Rees} M.~J., 1978, \mnras, 183, 341

\bibitem[{{Woosley} \& {Hoffman}(1992)}]{Woosley92}
{Woosley} S.~E., {Hoffman} R.~D., 1992, \apj, 395, 202

\bibitem[{{York} {et~al}\mbox{.}(2000){York}, {Adelman}, {Anderson},
  {Anderson}, {Annis}, {Bahcall}, {Bakken}, {Barkhouser}, {Bastian}, {Berman},
  {Boroski}, {Bracker}, {Briegel}, {Briggs}, {Brinkmann}, {Brunner}, {Burles},
  {Carey}, {Carr}, {Castander}, {Chen}, {Colestock}, {Connolly}, {Crocker},
  {Csabai}, {Czarapata}, {Davis}, {Doi}, {Dombeck}, {Eisenstein}, {Ellman},
  {Elms}, {Evans}, {Fan}, {Federwitz}, {Fiscelli}, {Friedman}, {Frieman},
  {Fukugita}, {Gillespie}, {Gunn}, {Gurbani}, {de Haas}, {Haldeman}, {Harris},
  {Hayes}, {Heckman}, {Hennessy}, {Hindsley}, {Holm}, {Holmgren}, {Huang},
  {Hull}, {Husby}, {Ichikawa}, {Ichikawa}, {Ivezi{\'c}}, {Kent}, {Kim},
  {Kinney}, {Klaene}, {Kleinman}, {Kleinman}, {Knapp}, {Korienek}, {Kron},
  {Kunszt}, {Lamb}, {Lee}, {Leger}, {Limmongkol}, {Lindenmeyer}, {Long},
  {Loomis}, {Loveday}, {Lucinio}, {Lupton}, {MacKinnon}, {Mannery}, {Mantsch},
  {Margon}, {McGehee}, {McKay}, {Meiksin}, {Merelli}, {Monet}, {Munn},
  {Narayanan}, {Nash}, {Neilsen}, {Neswold}, {Newberg}, {Nichol}, {Nicinski},
  {Nonino}, {Okada}, {Okamura}, {Ostriker}, {Owen}, {Pauls}, {Peoples},
  {Peterson}, {Petravick}, {Pier}, {Pope}, {Pordes}, {Prosapio},
  {Rechenmacher}, {Quinn}, {Richards}, {Richmond}, {Rivetta}, {Rockosi},
  {Ruthmansdorfer}, {Sand ford}, {Schlegel}, {Schneider}, {Sekiguchi},
  {Sergey}, {Shimasaku}, {Siegmund}, {Smee}, {Smith}, {Snedden}, {Stone},
  {Stoughton}, {Strauss}, {Stubbs}, {SubbaRao}, {Szalay}, {Szapudi}, {Szokoly},
  {Thakar}, {Tremonti}, {Tucker}, {Uomoto}, {Vanden Berk}, {Vogeley},
  {Waddell}, {Wang}, {Watanabe}, {Weinberg}, {Yanny}, {Yasuda}, \& {SDSS
  Collaboration}}]{York00}
{York} D.~G. {et~al.}, 2000, \aj, 120, 1579

\bibitem[{{Yuan} {et~al}\mbox{.}(2022){Yuan}, {Martin}, {Ibata}, {Caffau},
  {Bonifacio}, {Mashonkina}, {Errani}, {Doliva-Dolinsky}, {Starkenburg},
  {Venn}, {Arentsen}, {Aguado}, {Bellazzini}, {Famaey}, {Fouesneau},
  {Gonz{\'a}lez Hern{\'a}ndez}, {Jablonka}, {Lardo}, {Malhan}, {Navarro},
  {S{\'a}nchez Janssen}, {Sestito}, {Thomas}, {Viswanathan}, \&
  {Vitali}}]{Yuan22}
{Yuan} Z. {et~al.}, 2022, \mnras, 514, 1664

\bibitem[{{Yuan} {et~al}\mbox{.}(2025){Yuan}, {Matsuno}, {Sitnova}, {Martin},
  {Ibata}, {Ardern-Arentsen}, {Carlberg}, {Gonz{\'a}lez Hern{\'a}ndez},
  {Holmbeck}, {Kordopatis}, {Jiang}, {Malhan}, {Navarro}, {Sestito}, {Venn},
  {Viswanathan}, \& {Vitali}}]{Yuan25}
---, 2025, \aap, 698, A82

\bibitem[{{Zaremba} {et~al}\mbox{.}(2025){Zaremba}, {Venn}, {Hayes}, {Errani},
  {Cornejo}, {Glover}, {Jensen}, {McConnachie}, {Navarro}, {Pazder}, {Sestito},
  {Anthony}, {Andersen}, {Baker}, {Chin}, {Churilov}, {Diaz}, {Farrell},
  {Firpo}, {Gomez-Jimenez}, {Henderson}, {Kalari}, {Lawrence}, {Margheim},
  {Miller}, {Robertson}, {Ruiz-Carmona}, {Silversides}, {Silva}, {Young}, \&
  {Zhelem}}]{Zaremba25}
{Zaremba} D. {et~al.}, 2025, \apj, 987, 217

\end{thebibliography}

\label{lastpage}




\bsp	

\end{document}